\documentstyle[12pt,amsmath,amssymb]{article}
\def\Diag{{\hbox{\rm Diag}}}
\def\nuc{{\hbox{\scriptsize\rm UNC}}}
\def\SO{${\cal S}{\cal O}\,\,$}
\def\SOmath{{\cal S}{\cal O}}
\def\bR{{\mathbf{R}}}
\def\Dom{{\hbox{\rm Dom}}}
\def\Tr{{\hbox{\rm Tr}}}
\def\bS{{{\mathbf{S}}}}
\def\cR{{\cal R}}
\def\bC{{\mathbf{C}}}
\def\Det{{\hbox{\rm Det}}}
\def\e{{\rm e}}
\def\Fro{{\hbox{\scriptsize\rm Fro}}}
\def\Lin{{\hbox{\rm Lin}}}
\def\Aff{{\hbox{\rm Aff}}}
\def\det{{\hbox{\rm\scriptsize d}}}
\def\rand{{\hbox{\rm\scriptsize r}}}
\def\cF{{\cal F}}
\def\Card{{\hbox{\rm Card}}}
\def\Id{{\hbox{\rm Id}}}
\def\Col{\hbox{\rm Col}}

\def\bE{{\mathbf{E}}}
\def\bR{{\mathbf{R}}}
\def\cC{{\cal C}}
\def\cP{{\cal P}}
\def\Opt{{\hbox{\rm Opt}}}
\def\argmin{{\hbox{\rm argmin}}}
\def\Prox{{\hbox{\rm Prox}}}
\def\sign{{\hbox{\rm sign}}}
\def\SadVal{\hbox{\rm SadVal}}
\def\DualityGap{\hbox{\rm\small DualityGap}}
\def\Prob{\hbox{\rm Prob}}
\def\bV{{\mathbf{V}}}
\def\cL{{\cal L}}
\def\cA{{\cal A}}
\def\prox{{\hbox{\scriptsize\rm prox}}}
\def\nuc{{{\hbox{\scriptsize\rm nuc}}}}
\newcommand{\qed}{\hfill $\Box$ \par\noindent \medskip}
\newtheorem{theorem}{Theorem}[section]
\newtheorem{lemma}{Lemma}[section]
\newtheorem{proposition}{Proposition}[section]
\newtheorem{remark}{Remark}[section]
\begin{document}
\title{On solving large scale polynomial convex problems  by randomized first-order algorithms
\thanks{Research of both authors was supported by the BSF grant \# 2008302. Support of the second author was also supported by NSF grants DMS-0914785 and CMMI-1232623}}
\author{Aharon Ben-Tal \\
              Faculty of Industrial Engineering and Management, Technion\\
              Technion city, Haifa 32000, Israel \\
              {\tt abental@ie.technion.ac.il}
           \and
           Arkadi Nemirovski\\
              Georgia Institute of Technology,
              Atlanta, Georgia 30332, USA \\
              {\tt nemirovs@isye.gatech.edu}
}
\maketitle
\begin{abstract}
One of the most attractive recent approaches to processing well-structured large-scale convex optimization problems is based on smooth convex-concave saddle point reformulation of the problem of interest and solving the resulting problem by a fast First Order saddle point method utilizing smoothness of the saddle point cost function.
In this paper, we demonstrate that when the saddle point cost function is polynomial, the precise gradients of the cost function required by deterministic First Order saddle point algorithms and becoming prohibitively computationally expensive in the extremely large-scale case, can be replaced
with incomparably cheaper computationally unbiased random estimates of the gradients. We show that for large-scale problems with favourable  geometry, this randomization accelerates, progressively as the sizes of the problem grow, the solution process. This extends significantly previous results on acceleration by randomization, which, to the best of our knowledge, dealt solely with {\sl bilinear} saddle point problems. We illustrate our theoretical findings by instructive and  encouraging numerical experiments.
\end{abstract}
\section{Introduction}
The goal of this paper is to develop {\sl randomized} First Order algorithms for solving large-scale ``well structured'' convex-concave saddle point problems. The background and motivation for our work can be briefly outlined as follows. Theoretically, all of convex programming is within the grasp of polynomial time Interior Point Methods capable of generating high-accuracy solutions at a low iteration count. However, the complexity of an IPM iteration, in general, grows rapidly (as $n^3$) with the design dimension $n$ of the problem, which in numerous applications (like LP's with dense constraint matrices  arising in Signal Processing)  make IPM's prohibitively time-consuming in the large-scale case. There seemingly is consensus that ``beyond the practical grasp of IPM's,'' one should use the First Order Methods (FOM's) which, under favorable circumstances,   allow to get medium-accuracy solutions in (nearly) dimension-independent number of relatively cheap iterations. Over the last decade, there was a significant progress in FOM's; to the best of our understanding, the key to this progress is in discovering a way (Nesterov 2003, see \cite{smoothing}) to utilize problem's structure in order to accelerate FOM algorithms, specifically, to reduce a convex minimization problem $\min_{x\in X} f(x)$ with potentially nonsmooth objective $f$ to a saddle point problem
$$
\min_{x\in X}\max_{\sl y\in Y}\phi(x,y),\eqno{(SP)}
$$
 where $\phi$ is a  C$^{1,1}$  convex-concave function such that
\begin{equation}\label{ieq1}
f(x)=\max_{y\in Y}\phi(x,y).
\end{equation}
The  rationale is as follows: when $f$ is nonsmooth (which indeed is the case in typical applications), the (unimprovable in the large-scale case) rate of convergence of FOM's directly applied to the problem of interest $\min_{x\in X} f(x)$ is as low as $O(1/\sqrt{t})$, so that finding a feasible $\epsilon$-optimal solution takes as much as $O(1/\epsilon^2)$ iterations. Utilizing representation (\ref{ieq1}), this rate can be improved to $O(1/t)$; when $X$, $Y$ are simple, this  dramatic acceleration keeps the iteration's complexity basically intact.
\par
Now, in the original {\sl Nesterov's Smoothing} \cite{smoothing}, (\ref{ieq1}) is used to approximate $f$ by a C$^{1,1}$ function which is further minimized by Nesterov's optimal algorithm for smooth convex minimization originating from \cite{Yura83}. An alternative is to work on $(SP)$ ``as it is,'' by applying to $(SP)$ an $O(1/t)$-converging saddle point FOM, like the Mirror Prox algorithm \cite{DMP}; in what follows, we further develop this alternative.
\par
When solving $(SP)$ by a FOM, the computational effort per iteration has two components: (a) computing the values of $\nabla\phi$ at $O(1)$ points from $Z=X\times Y$, and  (b) ``computational overhead,'' like projecting onto $Z$. Depending on problem's structure and sizes, any one of these two components can become dominating; the approach we are developing in this paper is aimed at the situation where the computational ``expenses'' related to (a) by far dominate those related to (b), so that the ``practical grasp'' of the usual -- deterministic -- saddle point FOMs as applied to $(SP)$ is restricted with the problems where the required number of computations of $\nabla\phi$ (which usually is in the range of hundreds) can be carried out in a reasonable time. An attractive way to lift, to some extent, these restrictions is to pass from the precise values of $\nabla\phi$, which can be prohibitively costly computationally in the large-scale case, to computationally cheap unbiased {\sl random estimates} of these values. This idea (in retrospect, originating from the ad hoc sublinear type matrix game algorithm of Grigoriadis and Khachiyan \cite{GriKha}) has been developed in several papers, see \cite{Arora,NJLS,SMP,Baes,Rand}, \cite[section 6.5.2]{FOM} and references therein. To the best of our knowledge, for the time being ``acceleration via randomization'' was developed solely for the case of saddle point problems with {\sl bilinear} cost function $\phi$. The contribution of this paper is in extending the scope of randomization to the case of when $\phi$ is a {\sl polynomial}. \par
The main body of this paper is organized as follows. In section \ref{sect:SitGoals}, we formulate the problem of interest and present the necessary background on our ``working horse'' --- Mirror Prox algorithm. In section \ref{sect:rand}, we develop a general randomization scheme aimed at producing unbiased random estimates of $\nabla \phi$ for a polynomial $\phi$. Theoretical efficiency estimates for the resulting  randomized saddle point algorithm  are derived in section \ref{sect:complanal}. In section \ref{sectIll}, we illustrate our approach by working out in full details two generic examples: optimizing the maximal eigenvalue of a quadratic matrix pencil, and low dimensional approximation of a finite collection of points. We show theoretically (and illustrate by numerical examples) that in both these cases, in  a meaningful range of problem's sizes and $\epsilon$, solving problem within accuracy $\epsilon$ by randomized algorithm is by far less demanding computationally than achieving the same goal with the best known to us deterministic competitors, and the resulting ``acceleration by randomization'' goes to $\infty$ as the problem sizes grow. 
\section{Situation and Goals}\label{sect:SitGoals}
\subsection{Problem Statement}\label{sec:POI}
Consider the situation as follows: let $X\subset E_x,y\subset E_y$ be convex compact subsets of Euclidean spaces, and let $\phi(x,y): E:=E_x\times E_y\to\bR$  be a polynomial of degree $d$:
\begin{equation}\label{phinew}
\phi(\underbrace{x,y}_{z})=\sum_{k=0}^dQ_k(\underbrace{z,...,z}_{k}),
\end{equation}
where $Q_0$ is a constant, and for $k>0$, $Q_k(z^1,...,z^k)$ is a $k$-linear symmetric form on $E$ . From now on we assume that {\sl $\phi(x,y)$ is convex-concave on $X\times Y$}, that is, convex in $x\in X$ for fixed $y\in Y$, and concave in $y\in Y$ for fixed $x\in X$. Our problem of interest is the saddle point problem

\begin{equation}\label{SPP}
\SadVal=\min_{x\in X}\max_{y\in Y}\phi(x,y).
\end{equation}
Let
\begin{equation}\label{PD}
\begin{array}{rlcr}
\Opt(P)&=&\min\limits_{x\in X}\left[\overline{\phi}(x):=
\max_{y\in Y}\phi(x,y)\right]&(P)\\
\Opt(D)&=&\max\limits_{y\in Y}\left[\underline{\phi}(y):=
\min_{x\in X}\phi(x,y)\right]&(D)\\ \end{array}
\end{equation}
be the primal-dual pair of convex programs associated with (\ref{SPP}), so that $\Opt(P)=\Opt(D)$, let
\begin{equation}\label{DualGap}
\DualityGap(x,y)=[\overline{\phi}(x)-\Opt(P)]+[\Opt(D)-\underline{\phi}(y)]=\overline{\phi}(x)-\underline{\phi}(y)
\end{equation}
be the associated duality gap, and, finally, let
\begin{equation}\label{F}
F(z:=[x;y])=\left[F_x(x,y)=\phi^\prime_x(x,y);F_y(x,y):=-\phi^\prime_{y}(x,y)\right]:Z:=X\times Y\to E:=E_x\times E_y
\end{equation}
be the monotone mapping associated with (\ref{SPP}). Our ideal goal is, given tolerance $\epsilon>0$, to find an {\sl $\epsilon$-solution} to (\ref{SPP}), i.s., a point $z_\epsilon=(x_\epsilon,y_\epsilon)\in Z$ such that
\begin{equation}\label{goal}
\DualityGap(z_\epsilon)\leq\epsilon,
\end{equation}
whence $x_\epsilon$ is a feasible $\epsilon$-optimal solution to $(P)$, while $y_\epsilon$ is a feasible $\epsilon$-optimal solution to $(D)$.
We intend to achieve this goal by utilizing {\sl randomized} First Order saddle point algorithm, specifically, Stochastic Mirror Prox method (SMP) \cite{SMP}.
\subsection{Background on Stochastic Mirror Prox algorithm}\label{sec:SMP}
{\sl A setup} for SMP as applied to (\ref{SPP}) is given by
\begin{itemize}
\item a norm $\|\cdot\|$ on the subspace
$$L[Z]:=\Lin(Z-Z)$$
 in the embedding space $E:=E_x\times E_y$ of the domain $Z:=X\times Y$ of the saddle point problem. The (semi)norm on $E$ conjugate to $\|\cdot\|$ is denoted by $\|\cdot\|_*$:
    $$
    \|\zeta\|_*=\max\limits_{z\in L[Z]}\{\langle \zeta,z\rangle: \|z\|\leq1\};
    $$
\item a {\sl distance-generating function} (d.g.-f.) $\omega(z):Z\to\bR$ which should be convex and continuously differentiable on $Z$ and should be {\sl compatible} with $\|\cdot\|$, meaning strong convexity of $\omega(\cdot)$, modulus 1, w.r.t. $\|\cdot\|$:
    $$\langle \omega'(z)-\omega'(z'),z-z'\rangle \geq \|z-z'\|^2\,\,\forall (z,z'\in Z).$$
\end{itemize}
An SMP setup induces several important entities, specifically
\begin{itemize}
\item $\omega$-center $z_\omega:=\argmin_{z\in Z}\omega(z)$ of $Z$;
\item {\sl Bregman distance} $V_z(w):=\omega(w)-\omega(z)-\langle \omega'(z),w-z\rangle$, where $z,w\in Z$. By strong convexity of $\omega$, we have $V_z(w)\geq{1\over 2}\|w-z\|^2$;
\item {\sl $\omega$-radius} $\Omega:=\sqrt{2[\max_Z\omega(\cdot)-\min_Z\omega(\cdot)]}$; noting that ${1\over 2}\|w-z_\omega\|^2\leq V_{z_\omega}(w)\leq \omega(w)-\omega(z_\omega)$, we conclude that
\begin{equation}\label{raduis}
\forall (w\in Z): \|w-z\|\leq\Omega;
\end{equation}
\item {\sl Prox-mapping} $\Prox_z(\xi)$, $z\in Z$, $\xi\in E$, defined as
$$
\Prox_z(\xi)=\argmin_{w\in Z}\left[\langle \xi,w\rangle +V_z(w)\right]=\argmin_{w\in Z}\left[\langle \xi-\omega'(z),w\rangle +\omega(w)\right]
$$
\end{itemize}
As applied to (\ref{SPP}), SMP operates with {\sl Stochastic Oracle} representation of the vector field $F$ associated with the problem. A {\sl Stochastic Oracle} is a procedure (``black box'') which, at $t$-th call, a point $z_t$ being the input, returns the random vector
$$
g(z_t,\xi_t)=F(z_t)+\Delta(z_t,\xi_t)\in E
$$
where $\Delta(\cdot,\cdot)$ is a deterministic function, and $\xi_1,\xi_2,...$ is a sequence of i.i.d. ``oracle noises.'' The SMP algorithm is the recurrence
\begin{equation}\label{SMP}
\begin{array}{lrcl}\\
\hbox{initialization:}&z_1&=&z_\omega;\\
\hbox{search points:}&z_t&\mapsto&w_t=\Prox_{z_t}(\gamma_t g(z_t,\xi_{2t-1}))\mapsto z_{t+1}=\Prox_{z_t}(\gamma_t g(w_t,\xi_{2t}))\mapsto...\\
\hbox{approximate solutions:}&z^t&=&(x^t,y^t)=[\sum_{\tau=1}^t\gamma_\tau]^{-1}\sum_{\tau=1}^t\gamma_\tau w_\tau\\
\end{array}
\end{equation}
where $\gamma_t>0$ are deterministic stepsizes.\par
The main results on SMP we need are as follows (see the case $M=\mu=0$ of \cite[Corollary 1]{SMP}):
\begin{theorem}\label{theSMP} Assume that $\cL<\infty$ and $\sigma<\infty$ are such that
\begin{equation}\label{aresuch}
\begin{array}{ll}
(a)&\|F(z)-F(z')\|_*\leq \cL\|z-z'\|\,\,\forall z,z'\in Z\\
(b)&\bE_\xi\{\Delta(z,\xi)\}=0\,\,\forall z\in Z\\
(c)&\bE_\xi\{\|\Delta(z,\xi)\|_*^2\}\leq\sigma^2\,\,\forall z\in Z\\
\end{array}
\end{equation}
Then for every $t=1,2,...$ the $t$-step SMP with constant stepsizes
\begin{equation}\label{stepsizes}
\gamma_\tau=\min\left[{1\over\sqrt{3}\cL}, {\Omega\over\sqrt{7}\sigma\sqrt{t}}\right],\,1\leq\tau\leq t
\end{equation}
ensures that
\begin{equation}\label{mean}
\bE\{\DualityGap(x^t,y^t)\}\leq K_t:=\max\left[{2\Omega^2\cL\over t},{6\Omega\sigma\over\sqrt{t}}\right].
\end{equation}
In addition, strengthening {\rm (\ref{aresuch}.$b$,$c$)} to
\begin{equation}\label{strengthening}
\bE_\xi\{\Delta(z,\xi)\}=0,\,\bE\{\exp\{\|\Delta(z,\xi)\|_*^2/\sigma^2\}\}\leq\exp\{1\}
\end{equation}
we have an exponential bound on large deviations: for every $\Lambda>0$, we have
\begin{equation}\label{largedev}
\Prob\left\{\DualityGap(x^t,y^t)> K_t +\Lambda {7\Omega\sigma\over 2\sqrt{t}}\right\}\leq \exp\{-\Lambda^2/3\}+\exp\{-\Lambda t\}.
\end{equation}
\end{theorem}
\section{Randomization}\label{sect:rand}
Problem (\ref{SPP}) by itself is a fully deterministic  problem; with ``normal'' representation of the polynomial $\phi(x,y)$ (e.g., by the list of its nonzero coefficients), a precise ($\sigma=0$) deterministic oracle for $F$ is available; utilizing this oracle, a solution of accuracy $\epsilon$ is obtained in $O(1)\Omega^2\cL/\epsilon$ iterations, with  computational effort per iteration dominated by the necessity to compute the values of $F$ at two points and the values of two prox-mappings. When $Z$ is ``simple enough,'' the complexity of the second of these two tasks -- computing prox-mappings -- is a tiny fraction of the complexity of precise computation of the values of $F$. Whenever this is the case, it {\sl might} make sense to replace the precise values $F$ (which can be very costly in the large-scale case) with computationally cheap unbiased random estimates of these values. This is the option we intend to investigate in this paper. We start with a general description of the randomization we intend to use.
\par
Observe, first, that
$$
F(z)=D \nabla\phi (z)
$$
where $D=\Diag\{\Id_x,-\Id_y\}$, $\Id_x$ and $\Id_y$ being the identity mappings on $E_x$ and $E_y$, respectively.  Now, representing the polynomial $\phi(z)$ as
\begin{equation}\label{phi}
\phi(z)=\sum_{k=0}^dQ_k(\underbrace{z,...,z}_{k}),
\end{equation}
where $Q_k(z^1,...,z^k)$ is a symmetric $k$-linear form on $E$, differentiating (\ref{phi}) in $z$ and taking into account symmetry of $Q_k$, we have
\begin{equation}\label{formulaA}
\langle F(z),h\rangle =\langle D\nabla\phi(z),h\rangle=\langle \nabla\phi(z),Dh\rangle=\sum_{k=1}^d kQ_k(Dh,\underbrace{z,...,z}_{k-1})
\end{equation}
Now assume that we can associate with every $z\in Z$ a probability distribution $P_z$ on $E$ such that
\begin{equation}\label{NoBias}
\int \xi dP_z(\xi)=z\,\,\forall z\in E.
\end{equation}
In order to get an unbiased estimate of $F(z)$, one can act as follows:
\begin{itemize}
\item given $z$, draw $d-1$ independent samples $z^i\sim P_z$, $i=1,...,d-1$
\item compute the linear form $G=G[z^1,...,z^{d-1}]$ on $E$ given by
\begin{equation}\label{base}
\forall h\in E: \langle G,h\rangle=\sum_{k=1}^d kQ_k(Dh,z^1,z^2,...,z^{k-1}).
\end{equation}
thus ensuring that
\begin{equation}\label{unbiased}
\bE_{(z^1,...,z^{d-1})\sim P_z\times...\times P_z}\{G[z^1,...,z^{d-1}]\}=F(z)\,\,\forall z\in Z.
\end{equation}
\end{itemize}
Note that we can represent a random variable distributed according to $P_z$ as a deterministic function of $z$ and  random variable $\xi$ uniformly  distributed on $[0,1]$, which makes  $G$ a deterministic function of $z$ and $\xi\sim\hbox{Uniform}[0,1]$, as required by our model of a Stochastic Oracle.
\par
 Observe that for a general-type convex-concave polynomial $\phi(x,y)$ of degree $d$, precise deterministic computation of $F(z)$ is as suggested by (\ref{base})  {\sl with $P_z$ being the unit mass sitting at the singleton $z$}, that is, with $z^1=...=z^{d-1}=z$. It follows that {\sl if the distributions $P_z$, for every $z\in Z$ are such that computing the vectors $g_k$ of coefficients of the linear forms $Q_k(Dh,z^1,...,z^{k-1})$ of $h\in E$ is much cheaper than the similar task for the linear forms $Q_k(Dh,z,...,z)$ for a ``general position'' $z\in Z$, then computing the unbiased estimate $G=G[z^1,...,z^{d-1}]$  of $F(z)$ is much cheaper computationally than the precise computation of $F(z)$}, so that there are chances for the outlined randomization to reduce the overall complexity of computing $\epsilon$-solution to (\ref{SPP}). Let us look at a simple preliminary example:
\paragraph{Example 1} [``Scalar case'']: $E$ is just the space $\bR^n$ of $n$-dimensional vectors, and we have access to the coefficients of the $k$-linear forms $Q_k(\cdot)$ (e.g., $Q_k$ are given by lists of their   nonzero coefficients). In this case, we can specify $P_z$ as follows:
\par
Given $z\in E=\bR^n\backslash \{0\}$, let $P_z$ be the discrete probability distribution supported on the set $\{f^i=\sign(z_i)\|z\|_1e_i\}_{i=1}^n$, where $e^i$ are the standard basic orths in $E$, with the probability mass of $f^i$ equal to $|z_i|/\|z\|_1$; when $z=0$, let $P_z$ be the unit mass sitting at the origin. We clearly have $\bE_{f\sim P_z} \{f\}=z$, and all realizations of $f\sim P_z$ are extremely sparse --- with at most one nonzero entry. Now, in order to generate $f\sim P_z$, we need preprocessing of $O(1)n$ a.o. aimed to compute $\|z\|_1$ and the ``cumulative distribution''  $s_i=\|z\|_1^{-1}\sum_{j=1}^i|z_j|$, $i=1,...,n$. With this cumulative distribution at hand, to draw a sample $f\sim P_z$ takes just $O(1)\ln(n)$ a.o.:
we draw at random a real $\alpha$ uniformly distributed in $[0,1]$ (which for all practical purposed is just $O(1)$ a.o.), find by bisection the smallest $i\in\{1,...,n\}$ such that $\alpha\leq s_i$ ($O(1)\ln(n)$ a.o.) and return the index $i$ and the value $\sign(z_i)\|z\|_1$ of the only nonzero entry in the resulting vector $f$ ($O(1)$ a.o.). Thus, generating $z^1,...,z^{d-1}$ costs $O(1)[n+d\ln(n)]$ a.o. Now, with our ``ultimately sparse'' $z^1,...,z^{d-1}$, computing the $n$ coefficients of the linear form $Q_k(Dh,z^1,...,z^{k-1})$ of $h$ takes at most $O(1)[d+n\cC]$ a.o., where $\cC$ is an upper bound on the cost of extracting a coefficient of the $k$-linear symmetric form $Q_k$, $k\leq d$, given its ``address.'' The bottom line is that the complexity of computing $G[z^1,..,z^{d-1}]$ is
$$
\cC_r[P] =O(1)\left[n+d\ln(n)+d[d+n\cC]\right]=O(1)[d^2+dn\cC] \hbox{\ a.o.}
$$
On the other hand, computing $F(z)$ exactly costs something like
$$
\cC_d[P]=O(1)[n+\sum_{k=1}^d kN_k\cC]\hbox{\ a.o.}
$$
where $N_k$ is the total number of nonzero coefficients in $Q_k(\cdot,...,\cdot)$. Assuming that $d=O(1)$, we see that {\sl unless all $Q_k$ are pretty sparse -- just with $N_k=O(n)$ nonzero coefficients, mimicking unbiased Stochastic Oracle takes by orders of magnitude less computations than precise deterministic computation of $F(z)$.}
\section{Complexity Analysis}\label{sect:complanal}
The discussion in the previous section demonstrates that in some interesting cases unbiased random estimates of the vector field $F$ associated with (\ref{SPP}) are significantly cheaper computationally than the precise values of $F$. This does not mean, however, that in all these cases randomization is profitable --- it well may happen that as far as the overall complexity of $\epsilon$-solution is concerned, expensive high-quality local information is better than cheap low quality one. We intend to analyze the situation in the regime when the degree $d$ of the polynomial $\phi$ is a small integer formally treated as $O(1)$; this allows us to ignore in the sequel the details on how the hidden factors in $O(\cdot)$'s to follow depend on $d$.
\subsection{Preliminaries}
\paragraph{Standing Assumptions.} Observe that
\begin{equation}\label{direct1}
L[Z]:=\Lin(Z-Z)=\Lin(X-X)\times\Lin(Y-Y)=L[X]\times L[Y].
\end{equation}
 Now, the sets
$$ X^s={1\over 2}[X-X],\, Y^s={1\over 2}[Y-Y],\, Z^s={1\over 2}[Z-Z]=X^s\times Y^s$$
are unit balls of certain norms $\|\cdot\|_X$ on $L[X]$, $\|\cdot\|_Y$ on $L[Y]$ and $\|\cdot\|$ on $L[Z]$, with
\begin{equation}\label{direct2}
\|(x,y)\|=\max[\|x\|_X,\|y\|_Y],\,x\in L[X], y\in L[Y].
\end{equation}
From now on, we make the following
\begin{quote}
{\bf Assumption A.} {\sl The just defined norm $\|\cdot\|$ with the unit ball ${1\over 2}[Z-Z]$ is the norm used in the SMP setup, while the d.-g.f $\omega(x,y)$ is of the form $\omega_X(x)+\omega_Y(y)$, where $(\|\cdot\|_X,\omega_X(\cdot))$ and $(\|\cdot\|_Y,\omega_Y(\cdot))$ form SMP setups for $(X,E_x)$ and $(Y,E_y)$ respectively\footnote{Note that such a sum indeed is a d.-g.f. fitting the norm $\|\cdot\|$.}.}
 \end{quote}
 Note that
 \begin{itemize}
 \item We have
 \begin{equation}\label{direct3}
 \|[\xi;\eta]\|_*=\|\xi\|_{X,*}+\|\eta\|_{Y,*},
 \end{equation}
 where $\|\cdot\|_{X,*}$ and $\|\cdot\|_{Y,*}$ are the (semi)norms conjugate to $\|\cdot\|_X$, $\|\cdot\|_Y$, respectively. In particular, we have
\begin{equation}\label{unitary}
\|F(z)\|_*=\|\nabla\phi(z)\|_*,\,\,\|F(z)-F(z')\|_*=\|\nabla\phi(z)-\nabla\phi(z')\|_*\,\,\forall z,z'\in E.
\end{equation}
\item The $\omega$-radius $\Omega$ of $Z$ is
\begin{equation}\label{Omega}
\Omega=\sqrt{\Omega_X^2+\Omega_Y^2},\Omega_X=\sqrt{2[\max_{x\in X}\omega_X(x)-\min_{x\in X}\omega_X(x)]},\Omega_Y=\sqrt{2[\max_{y\in Y}\omega_Y(y)-\min_{y\in Y}\omega_Y(y)]}
\end{equation}
\end{itemize}

\paragraph{Scale factor.}  When speaking about complexity of finding $\epsilon$-solution, we shall express it in terms of the {\sl relative accuracy}
$\nu=\epsilon/\bV$, where the {\sl scale factor} $\bV$ is defined as follows. Let $\widehat{Z}$ be the convex hull of $\{0\}\cup Z$, and let
$$
\widehat{\phi}(z)=\phi(z)-\phi(0)-\langle \phi'(0),z\rangle=\sum_{k=2}^dQ_k(z,...,z).
$$
We set
\begin{equation}\label{V}
\bV=\bV_Z[\phi]:={\max}_{z\in \widehat{Z}} \widehat{\phi}(z)-{\min}_{z\in \widehat{Z}} \widehat{\phi}(z).
\end{equation}
The importance of this scale factor in our contents stems from the following simple observation (see also Lemma \ref{lemraash} below):
\begin{lemma}\label{lem1}
For properly chosen positive real $C^{(1)}$  depending solely on $d$, for all $k$, $2\leq k\leq d$ and all collections $z^1,...,z^k$ of vectors from $L[\widehat{Z}]$ one has
\begin{equation}\label{eq1}
|Q_k(z^1,...,z^k)|\leq C^{(1)}\bV\prod\limits_{i=1}^k\|z^i\|_{\widehat{Z}}
\end{equation}
In particular, the vector field $F(z)$ associated with {\rm (\ref{SPP})} satisfies {\rm (\ref{aresuch}.$a$)} with
\begin{equation}\label{cLis}
\cL=C^{(1)}\bV\sum_{k=2}^dk(k-1)2^{k-2}:=C^{(2)}\bV,
\end{equation}
where $C^{(2)}$ depends solely on $d$.
\end{lemma}
For proof, see Appendix.
\par
An immediate question related to the definition of the scaling factor is:  a ``shift of the problem by $a\in E$'' -- a simple substitution of variables $z=w-a$ -- changes the factor and thus the complexity estimates, although such a substitution leaves the problem ``the same.'' The answer is as follows: while the ``shift option'' should be kept in mind, such a shift changes the Stochastic Oracle  as given by (\ref{base}). Indeed, this oracle is defined in terms of the {\sl homogeneous components in the Taylor decomposition of $\phi(\cdot)$ taken at the origin}, and this is why the origin is participating in the description of $\widehat{Z}$ and thus of $\bV$. Shifting the origin, we, in general, change the \SO \footnote{For example, with $\phi(x,y)\equiv x^3$, the oracle (\ref{base}) is $G=[3x^1x^2;0]$, $z^i=[x^i;0]\sim P_z$. Substituting $x=1+h$, carrying out the construction of the \SO ``in $h$-variable'' and translating the result back to $x$-variable, the resulting \SO turns out to be $G=[3x^1x^2+3x^1-3x^2;0]$, which is not the  oracle we started with.}, and thus there is nothing strange that our scaling of the accuracy (and thus -- the efficiency estimates) corresponding to a given $Z$ and a given (implicitly participating in (\ref{base})) \SO is not translation-invariant.
\subsection{Complexity Analysis}
\paragraph{Preliminaries.}
From now on we assume that as applied to (\ref{SPP}), SMP utilizes Stochastic Oracle \SO  given according to (\ref{base}) by a family of probability distributions $\cP=\{P_z:z\in Z\}$ on $E$ satisfying (\ref{NoBias}). From now on, we make the following
\begin{quote}
{\bf Assumption B.} {\sl For some $\rho\geq 0$, all distributions $P_z$, $z\in Z$, are supported on the set $Z+2\rho Z^s\subset\Aff(Z)$, where $Z^s={1\over 2}[Z-Z]$ and $\Aff(Z)$ is the affine hull of $Z$.}
\end{quote}
In particular, when $P_z$ is supported on $Z$ for all $z\in Z$ (''proper case''), Assumption B is satisfied with $\rho=0$.
\par
It is time now to note that the \SO we have developed so far gives rise to a {\sl parametric family} of Stochastic Oracles, specifically, as follows. First of all, our basic \SO in fact can be ``split'' into two Stochastic Oracles, $\SOmath^x$ and $\SOmath^y$, providing estimates of the $x$- and the $y$-components $F_x,F_y$ of $F(z)=[F_x(z);F_y(z)]$:
the estimates
$$
\begin{array}{rcl}
E_x\ni G_x&=&G_x[z^1,...,z^{d-1}]: \forall \xi\in E_x: \langle G_x,\xi\rangle=\sum_{k=1}^dkQ_k([\xi;0],z^1,...,z^{k-1}),\\
E_y\ni G_y&=&G_y[z^1,...,z^{d-1}]: \forall \eta\in E_y: \langle G_y,\eta\rangle=-\sum_{k=1}^dkQ_k([0;\eta],z^1,...,z^{k-1}).\\
\end{array}
$$
Here, as above, $z^1,...,z^{d-1}$ are, independently of each other, sampled from $P_z$. Now, given two positive integers $k_x,k_y$, we can ``recombine'' our ``partial stochastic oracles''  $\SOmath^x$, $\SOmath^y$ into a new Stochastic Oracle $\SOmath_{k_x,k_y}$ as follows: in order to generate a random estimate of $F(z)$ given $z\in Z$, we generate $(d-1)\max[k_x,k_y]$ independent samples $z^k_\tau\sim P_z$, $1\leq k\leq d-1$, $1\leq\tau\leq k_{xy}:=\max[k_x,k_y]$ and then set
\begin{equation}\label{g}
g=G^{k_x,k_y}_z\left[\{z^k_\tau\}_{{1\leq k\leq d-1,\atop1\leq\tau\leq k_{xy}}}\right]=
\left[{1\over k_x}\sum_{\tau=1}^{k_x}G_x[z^1_\tau,...,z^{d-1}_\tau];{1\over k_y}\sum_{\tau=1}^{k_y}G_y[z^1_\tau,...,z^{d-1}_\tau]\right].
\end{equation}
In the sequel, we refer to $k_x$ and $k_y$ as the {\sl $x$- and $y$- multiplicities} of the Stochastic Oracle $\SOmath_{k_x,k_y}$.\par
We will make use of the following
\begin{lemma}\label{lemraash}
Under Assumptions A, B, for all positive integer multiplicities $k_x$, $k_y$, $\SOmath_{k_x,k_y}$ ensures validity of {\rm (\ref{aresuch}.$b$)}, same as the validity of
{\rm (\ref{strengthening})} with
\begin{equation}\label{sigmais}
\sigma=C^{(3)}\bV(1+\rho)^{d-1}\left[\min[1,\Omega_X/\sqrt{k_x}]+\min[1,\Omega_Y/\sqrt{k_y}]\right],
\end{equation}
where C$^{(3)}$ depends solely on $d$.
\end{lemma}
For proof, see Appendix.
\par
We have arrived at the following
\begin{theorem}\label{themain} Let $t\geq 1$ be given, let Assumptions A, B be satisfied, and let problem {\rm (\ref{SPP})} be solved by $t$-step SMP utilizing $\SOmath_{k_x,k_y}$, with the parameters $\cL$, $\sigma$ underlying the stepsize policy {\rm (\ref{stepsizes})} given by {\rm (\ref{cLis}), (\ref{sigmais})}. Then, for some $C$ depending solely on $d$,
\begin{equation}\label{totalb}
\begin{array}{ll}
(a)&\bE\{\DualityGap(x^t,y^t)\}\leq K(t):=C\left[{\Omega_X^2+\Omega_Y^2\over t}+{\sqrt{\Omega_X^2+\Omega_Y^2}(1+\rho)^{d-1}\vartheta\over\sqrt{t}}\right]\bV\\
&\multicolumn{1}{r}{\vartheta=\min[1,\Omega_X/\sqrt{k_x}]+\min[1,\Omega_Y/\sqrt{k_y}];}\\
(b)&\Prob\left\{\DualityGap(x^t,y^t)>K(t)+C\Lambda{\sqrt{\Omega_X^2+\Omega_Y^2}(1+\rho)^{d-1}\vartheta\bV\over\sqrt{t}}\right\}\leq\exp\{-\Lambda^2/3\}+\exp\{-\Lambda t\}.\\
&\multicolumn{1}{r}{\forall \Lambda>0.}\\
\end{array}
\end{equation}
\end{theorem}
\section{Illustrations}\label{sectIll}
We illustrate the proposed approach by two examples. The first of them is of a purely academic nature, the second can pretend to be of some applied interest. When selecting the examples, our major goal was to illustrate randomization schemes different  from the one in Example 1.
\subsection{Illustration I: minimizing the maximal eigenvalue of a quadratic matrix pencil}\label{Ill1}
\paragraph{The problem} we are interested in is as follows: We are given a symmetric matrix quadratically depending on the ``design variables'' $x_1,...,x_J$ which themselves are matrices:
\begin{equation}\label{cA}
\cA(x)=\sum_{i=1}^I \left[a_i^Tx_{j(i)}^Tq_ix_{j(i)}a_i+b_i^Tx_{j(i)}c_i+c_i^Tx_{j(i)}^Tb_i\right]+d\in E_y:=\bS^m,
\end{equation}
where
\begin{itemize}
\item $\bS^m$ is the space of $m\times m$ symmetric matrices equipped with the Frobenius inner product,
\item $x=\{x_j\in\bR^{m_j\times n_j}\}_{j=1}^J$ is a collection of variable matrices which we treat as a block-diagonal rectangular matrix with diagonal blocks $x_j$, $1\leq j\leq J$. We denote the linear space of all these matrices by $E_x$ and equip it with the Frobenius inner product;
\item $j(i)\in\{1,...,J\}$, $1\leq i\leq I$, are given integers,
\item $\{a_i,b_i,c_i,q_i\}_{i=1}^I$, $d$ are data matrices of appropriate sizes and structures:
$$
a_i,c_i\in\bR^{n_{j(i)}\times m},\,b_i\in\bR^{m_{j(i)}\times m},\,q_i\in\bS^{m_{j(i)}},\,d\in\bS^m;
$$
in addition, we assume that {\sl all $q_i$ are positive semidefinite}, and that the values $j(i)$, $1\leq i\leq I$, cover the entire range $1\leq j\leq J$, meaning that every one of the blocks $x_j$ indeed participates in $\cA(\cdot)$.
\end{itemize}
For a matrix $a\in\bR^{p\times q}$, let $\sigma(a)=[\sigma_1(a);...;\sigma_{\min[p,q]}(a)]$ be the vector of singular values of $a$ arranged in the non-ascending order, and let $\|a\|_\nuc=\|\sigma(a)\|_1$ be the nuclear norm of $a$. For a symmetric matrix $a$, let $\lambda_{\max}(a)$ be the maximal eigenvalue of $a$. Finally, let $$X=\{x\in E_x:\|x\|_\nuc\leq1\}.$$ Our goal is to solve the optimization problem
\begin{equation}\label{prb}
\Opt=\min_{x\in X} \left\{\lambda_{\max}(\cA(x))\right\},
\end{equation}
Denoting by $Y$ the standard {\sl spectahedron} in $\bS^m$:
$$
Y=\{y\in\bS^m:y\succeq0, \Tr(y)=1\}
$$
and observing that $\lambda_{\max}(a)=\max_y\{\Tr(ay):y\in Y\}$, we can convert the problem of interest into the saddle point problem as follows:
\begin{equation}\label{SPI}
\Opt=\min_{x\in X}\max_{y\in Y}\left[\phi(x,y):=\Tr(y\cA(x))\right].
\end{equation}
From $q_i\succeq0$, $i\leq I$, and the fact that $y\succeq0$ for all $y\in Y$ it immediately follows that the restriction of $\phi$ on $y\in Y$ is convex in $x\in E_x$; as a function of $y$, $\phi$ is just linear. Thus, $\phi$ is a convex-concave on $X\times Y$ polynomial of degree $d=3$. The monotone mapping (\ref{F}) associated with (\ref{SPI}) is
\begin{equation}\label{FI}
\begin{array}{rcl}
F_x(x,y)&=&2\Diag\{\sum_{i:j(i)=j}[q_ix_ja_iya_i^T+b_iyc_i^T],1\leq j\leq J\}\in E_x,\\
F_y(x,y)&=&-\cA(x),\\
\end{array}
\end{equation}
Now let us apply to (\ref{SPI}) the approach we have developed so far.
\par
{\bf A.} First, let us fix the setup for SMP. We are in the situation when $X^s:={1\over 2}[X-X]$ is $X$ -- the unit ball of the nuclear norm on $E_x$; thus, $\|\cdot\|_X$ is the nuclear norm on $E_x$. The set $Y^s={1\over 2}[Y-Y]$ clearly is contained in the unit nuclear norm ball of $\bS^m$ and contains the concentric nuclear norm ball of radius $1/2$, meaning that $\|\cdot\|_Y$ is within factor 2 of the nuclear norm:
$$
2\|y\|_\nuc\geq \|y\|_Y\geq \|y\|_\nuc\,\,\forall y\in \bS^m=E_y.
$$
The best, within $O(1)$ factors, known so far under circumstances choice of the d.-g.f.'s is (see \cite[section 5.7.1]{FOM} or Propositions \ref{propNN2}, \ref{propNN1} in Appendix)
\begin{equation}\label{dgfs}
\begin{array}{rcl}
\omega_X(x=\Diag\{x_1,...,x_J\})&=&O(1)\ln(n)\sum_{j=1}^J\sum_{\ell=1}^{\min[m_j,n_j]}\sigma_\ell^{q({n})}(x_j),\\
&&{n}=\sum_{j=1}^J\min[m_j,n_j],\,q(n)={1\over 2\ln({n})},\\
 \omega_Y(y)&=&O(1)\ln(m)\sum_{\ell=1}^m \sigma_\ell^{q(m)}(y),\\
 \end{array}\ \
 \footnotemark^)
 \end{equation}
\footnotetext{To avoid trivial situations, we assume from now on that $m>1$, $n>1$.}\noindent
with explicitly given absolute constants $O(1)$. This choice is reasonably good in terms of the values of the corresponding radii of $X$, $Y$ which turn to be ``quite moderate:''
\begin{equation}\label{radii}
\Omega_X\leq O(1)\sqrt{\ln({n})},\,\,\Omega_Y\leq O(1)\sqrt{\ln(m)}.
\end{equation}
Note that the efficiency estimate (\ref{totalb}) says that we are interested in as small values of $\Omega_X$, $\Omega_Y$ as possible. At the same time, it is immediately seen that if $\omega(\cdot)$ is a d.-g-.f. for $Z$ compatible with the norm generated by $Z$ (i.e., with the unit ball $Z^s={1\over 2}[Z-Z]$, then the $\omega$-radius of $Z$ is {\sl at least} $O(1)$, so that $\Omega_X$, $\Omega_Y$ are ``nearly as good'' as the could be.
\par
The outlined d.-g.f.'s are also the best known under circumstances in terms of the computational complexity of the associated prox-mapping; it is easily seen that this complexity is dominated by the necessity to carry out singular value decomposition of a matrix from $E_x$ (which takes $O(\sum_jm_j n_j\min[m_j,n_j])$ a.o.) and eigenvalue decomposition of a matrix from $\bS^m$ ($O(m^3)$ a.o.), see below.
\par
{\bf B.} With our approach, the ``basic'' option when solving (\ref{SPI}) is to use the deterministic version of SMP, i.e., to use as $P_z$ the unit mass sitting at $z$. The corresponding efficiency estimate can be obtained from (\ref{totalb}) by setting $k_x=k_y=\infty$; taking into account (\ref{radii}), the resulting estimate says that a solution to (\ref{SPI}) of a given accuracy $\epsilon\leq\bV$ will be found in course of
\begin{equation}\label{Nofeps}
N_\det(\epsilon/\bV)=O(1)\ln(m{n}){\bV/\epsilon}
\end{equation}
iterations. Now let us evaluate the arithmetic complexity of an iteration. From the description of the algorithm it is clear than the computational effort at an iteration is dominated by the necessity to compute exactly $O(1)$ values of the monotone mapping (\ref{FI}) and of $O(1)$ prox mappings. To simplify evaluating the computational cost of an iteration, assume from now on that we are in the {\sl simple case}:
$$
 m_j=n_j=\nu,\,1\leq j\leq J.
$$
In this case, computing $O(1)$ values of the prox mapping costs
$$
\cC_\prox=O(1)[m^3+J\nu^3]\hbox{\ a.o.}
$$
\begin{quote}
{\small Indeed, with our $\omega_X(\cdot)$, computing the $x$-component of  prox mapping reduces to solving the optimization problem $\min_{v\in E_x,\|v\|_\nuc\leq1}[\sum_{j=}^J\sum_{\ell=1}^\nu\sigma_\ell^q(v_j) -\Tr(g^Tv)]$ with a given $q\in(1,2]$ and a given $g\in E_x$. To solve the problem, we compute the singular value decompositions of all diagonal blocks $g_j$ in $g$, this getting a representation $g=U\Diag\{\gamma\}V^T$ with block-diagonal orthogonal matrices $U$, $V$, which takes $O(1)J\nu^3$ a.o. It is immediately seen that the problem admits an optimal solution $v$ of the same structure as $g$: $v=U\Diag\{\upsilon\}V^T$. Specifying $\upsilon$ reduces to solving the convex optimization problem $$\min_{\upsilon\in\bR^n:\|\upsilon\|_1\leq1}\left[\sum_j[|\upsilon_j|^q+\gamma_j\upsilon_j]\right];$$
 this convex problem with separable objective and a single separable constraint clearly can be solved within machine precision in $O(n)$ a.o. Finally, given $\upsilon$, it takes $O(1)J\nu^3$ operations to compute the $x$-component $U\Diag\{\upsilon\}V^T$ of the prox mapping. Thus, the total cost of the $x$-component of the prox mapping is $O(1)J\nu^3$ a.o. The situation with computing the $y$-component of the mapping is completely similar, and the cost of this component is $O(1)m^3$ a.o.}
\end{quote}
Looking at (\ref{FI}), we see that computing $O(1)$ values of $F$ at  ``general position'' points $z$, assuming all the data matrices dense, is
$$
\cC_F=O(1)\nu m (\nu+m) I \hbox{\ a.o.}
$$
As a result, the arithmetic cost of finding $\epsilon$-solution to (\ref{SPI}) (and thus -- to (\ref{prb})) by the deterministic version of SMP is
\begin{equation}\label{complIdet}
\cC_\det (\epsilon)=O(1)\ln(mn)\underbrace{\big[\overbrace{m^3+ J\nu^3}^{\Theta_\prox}+\overbrace{m\nu (m+\nu) I}^{\Theta_F}\big]}_{\Theta}{\bV\over\epsilon}\hbox{\ a.o.}
\end{equation}
Note that we are not aware of better complexity bounds for large-scale problems (\ref{prb}), at least in the  case when in the expression for $\Theta$, the term $m^3$ is dominated by the sum of other terms.
\par
{\bf C.} Now let us look whether we can reduce the overall arithmetic cost of $\epsilon$-solution to (\ref{prb}) by randomization. An immediate observation  is that the only case when it can happen is the one of $\Theta_F\gg\Theta_\prox$. Indeed, comparing the efficiency estimates (\ref{totalb}) and (\ref{Nofeps}), we conclude that randomization can only increase the iteration cost of $\epsilon$-solution; in order to overweigh the growth in the number of iterations, we need to reduce significantly the arithmetic cost of an iteration, and to this end, this cost, in the deterministic case, should be by far dominated by the cost of computing the values of $F$ (the only component of our computational effort which can be reduced by randomization). Assuming $\Theta_F\gg \Theta_\prox$, let us look which kind of randomization could be useful in our context. Note that in order for randomization to be useful, the underlying distributions $P_z$ should be supported on the set of those pairs $(x,y)$ for which computing an estimate $g$ of $F(z)$ according to (\ref{g}) is much cheaper than computing $F$  at a general-type point $(x,y)\in E_x\times E_y$. A natural way to meet this requirement us to use the ``matrix analogy'' of Example 1, where $P_z$ are supported on the set of low rank matrices. Specifically, in order to get an unbiased estimate of $F(z)$, $z=(x,y)\in X\times Y$, let us act as follows:
\begin{enumerate}
\item We compute the singular value decomposition $x=U\Diag\{\sigma[x]\}V^T$ of $x= \Diag\{x_1,...,x_J\}$ (here $\sigma[x]=[\sigma(x_1);...;\sigma(x_J)]$) and eigenvalue decomposition $y=W\Diag\{\sigma(y)\}W^T$ of $y$ \footnote{Note that the singular values of $y$ are the same as eigenvalues, since $y\succeq0$  due to $y\in Y$.}, where $U,V$ are block-diagonal $n\times n$ orthogonal with $\nu\times\nu$ diagonal blocks, and $W$ is an orthogonal $m\times m$ matrix.
\item We specify $P_x$ as the distribution of a random matrix $\xi\in E_x$ with takes the values $$\|\sigma[x]\|_1\Col_\ell[U]\Col_\ell^T[V],\, 1\leq \ell\leq n,$$ with the probabilities $(\sigma[x])_\ell/\|\sigma[x]\|_1$ (when $\sigma[x]=0$, $\xi$ takes value $0$ with probability 1); here $\Col_\ell(A)$ denotes $\ell$-th column of  a matrix $A$.
\item We specify $P_y$ as the distribution of the random symmetric matrix $\eta$ which takes values $\Col_i[W]\Col_i^T[W]$, $1\leq i\leq m$, with probabilities $\sigma_i(y)$, and specify $P_z$ as the direct product of $P_x$ and $P_y$.
\end{enumerate}
Observe that the expectation of $\zeta\sim P_z$ is exactly $z$, and that $P_z$, $z\in Z=X\times Y$,  is supported on $Z$ due to $\|\sigma(x)\|_1\leq1$, $x\in X$, $\|\sigma(y)\|_1=1$, $y\in Y$. In other words, {\sl assumption B is satisfied with $\rho=0$.}
\par
Note that with the just defined $P_z$, a realization $\zeta=(\xi,\eta)\sim P_z$ is of very special structure:
\begin{equation}\label{structure}
\xi=u\times v^T,\,u,v\in\bR^n,\,\,\eta=ww^T,\,w\in\bR^m;
\end{equation}
moreover, among the $J$ consecutive $\nu$-dimensional blocks $u_j,v_j$, $j=1,...,J$, of every one of the vectors $u,v\in\bR^{n=J\nu}$, all but one blocks are zero, and the nonzero blocks $u_j$, $v_j$ share a common index $j$.
\par
It is immediately seen that with the just defined distributions $P_z$, the unbiased estimate (\ref{g}) of $F(z)$ is as follows:
\begin{equation}\label{gI}
\begin{array}{rcl}
G_x^{k_x}&=&{1\over k_x}\sum\limits_{\ell=1}^{k_x}\Diag\bigg\{\sum\limits_{i:j(i)=j}\big[q_i u^{2\ell-1}_j[v^{2\ell-1}_j]^Ta_iw^{2\ell}[w^{2\ell}]^Ta_i^T
+q_i u^{2\ell}_j[v^{2\ell}_j]^Ta_iw^{2\ell-1}[w^{2\ell-1}]^Ta_i^T\\
&&\qquad\qquad\qquad\qquad\qquad\qquad\qquad\qquad\qquad\qquad+2b_iw^{2\ell-1}[w^{2\ell-1}]^Tc_i^T\big],j=1,...,J\bigg\}\\
G_y^{k_y}&=&-d-{1\over k_y}\sum\limits_{\ell=1}^{k_y}\sum\limits_{i=1}^I\bigg[{1\over 2}a_i^Tv^{2\ell-1}_{j(i)}[u^{2\ell-1}_{j(i)}]^Tq_iu^{2\ell}_{j(i)}[v^{2\ell}_{j(i)}]^Ta_i+{1\over 2}a_i^Tv^{2\ell}_{j(i)}[u^{2\ell}_{j(i)}]^Tq_iu^{2\ell-1}_{j(i)}[v^{2\ell-1}_{j(i)}]^T\\
&&\qquad\qquad\qquad\qquad\qquad\qquad\qquad\qquad\qquad
+b_i^Tu^{2\ell-1}_{j(i)}[v^{2\ell-1}_{j(i)}]^Tc_i+c_i^Tv^{2\ell-1}_{j(i)}[u^{2\ell-1}_{j(i)}]^Tb_i\bigg]\\
\end{array}
\end{equation}
where the collections
$$\zeta^\ell=\left([u^\ell_1;...;u^\ell_J][v^\ell_1;...;v^\ell_J]^T,w^\ell[w^\ell]^T\right),\,\ell=1,...,2\max[k_x,k_y]$$
are independently of each other drawn from $P_z$.
\par
It is immediately seen that the arithmetic cost of computing $(G_x,G_y)$ given $z=(x,y)$ is comprised of the components as follows:
\begin{enumerate}
\item ``Setup cost'' -- one of computing singular value decomposition of $x$ and eigenvalue decomposition of $y$ \footnote{In fact, this cost is nonexisting: by construction of the method, the points $z$ where one needs to evaluate $F$ are the values of already computed prox-mappings; according to how we compute these values (see above), they go together with their singular value/eigenvalue decompositions.} ($O(1)(m^3+J\nu^3)$ a.o.) plus the cost of computing the ``cumulative distributions'' $S_j(x)=\|\sigma[x]\|_1^{-1}\sum_{\tau=1}^j(\sigma[x])_\tau$, $1\leq j\leq J\nu$, $S_i(y)=\sum_{\tau=1}^i\sigma_\tau(y)$ ($O(1)(m+J\nu)$ a.o.).
\item After the setup cost is paid, for every $\ell$\\
--- generating $\zeta^\ell$ costs $O(1)(\ln(m)+\ln(J\nu)+m+\nu)$ a.o.,\\
--- computing the contribution of $(\zeta^{2\ell-1},\zeta^{2\ell})$ to $G_x$ costs no more than $O(1)I\nu(m+\nu)$ a.o. (look at (\ref{gI})), and this cost should be paid $k_x$ times;\\
--- computing the contribution of $(\zeta^{2\ell-1},\zeta^{2\ell})$ to $G_y$ costs at most $O(1)(m+\nu)^2K$ a.o., where $K=\max_{1\leq j\leq J} \Card\{i: j(i)=j\}$ (look at (\ref{gI}) and take into account that the vectors $u^{\ell}$, $v^{\ell}$ have a single nonzero $\nu$-dimensional block each), and this cost should be paid  $k_y$ times.
\end{enumerate}
Thus, the cost of computing $(G_x^{k_x},G_y^{k_y})$ is
\begin{equation}\label{costofG}
\begin{array}{c}
O(1)\left(m^3+J\nu^3+k_x \nu(m+\nu)I+ k_y (m+\nu)^2K+k_y\ln(J)\right)\hbox{\ a.o.},\\
K=\max_{1\leq j\leq J} \Card\{i: j(i)=j\}\\
\end{array}
\end{equation}
(note that $J\leq I$).
To simplify the analysis to follow, assume from now on that $I=J\leq\exp\{(m+\nu)^2K\}$ and $j(\cdot)$ is one-to-one. In this case $K=1$ and the cost of an iteration is
\begin{equation}\label{costrand}
O(1)\left(m^3+J\nu^3 + k_x\nu(m+\nu)J+k_y(m+\nu)^2\right)\hbox{\ a.o.}
\end{equation}
Now let us evaluate the overall complexity of finding, with confidence $1-\delta$, $\delta\ll1$, an $\epsilon$-solution by the randomized SMP. We assume from now on that $\epsilon\leq\bV$ (otherwise the problem is trivial, since $\DualityGap(z)\leq\bV$ for every $z\in X\times Y$). For the sake of simplicity, we restrict ourselves with the case of $k_x=k_y=1$. Invoking the efficiency estimate (\ref{totalb}.$b$) with $\Lambda=O(1)\sqrt{\ln(1/\delta)}$ and taking into account (\ref{radii}) and the fact that we are in the situation of $\rho=0$, the number $t$ of iterations which results,with confidence $1-\delta$, in $\DualityGap(x^t,y^t)\leq\epsilon$ is bounded from above by
$$
N_{\rand,\delta}(\epsilon)=O(1)\ln(mn)\ln(1/\delta)(\bV/\epsilon)^2,
$$
provided that $\ln(1/\delta)\leq O(1)\ln(mn)(\bV/\epsilon)^2$. Thus,
the iteration count now is nearly square of the one for the deterministic algorithm, see (\ref{Nofeps}). Taking into account  (\ref{costrand}), the overall complexity of achieving our goal with the randomized algorithm  does not exceed
$$
\cC_{\rand,\delta}(\epsilon)=O(1)\ln(mn)\ln(1/\delta)\left[m^3+J\nu^3+(m+\nu)(m+\nu J)\right](\bV/\epsilon)^2 \hbox{\ a.o.}
$$
The ratio of this quantity and the ``deterministic complexity'' (see (\ref{complIdet}) and take into account that we are in the case of $I=J$) is
$$
\cR={\cC_{\rand,\delta}(\epsilon)\over\cC_\det(\epsilon)}=O(1)\ln(1/\delta)\underbrace{\left[{m^3+\nu^3J+(m+\nu)(m+\nu J)\over m^3+\nu^3J+m\nu(m+\nu)J}\right]}_{r}\cdot{\bV\over\epsilon}.
$$
It is immediately seen that {\sl when $\bV/\epsilon$ and $\delta$ are fixed, and $m,\nu,J$ vary in such a way that $m,n=\nu J$ go to $\infty$ and $\nu/m$, $m/n$  go to 0}, $\cR$ goes to 0 as $O(1/m)$, meaning that eventually the randomized algorithm outperforms its deterministic competitor, and the ``performance ratio'' goes to $\infty$ as the sizes $m,n$ of the problem grow.
\paragraph{Numerical illustration.}  In the experiment we are about to describe, the sizes of problem (\ref{prb}) were selected as
$$
m=300,\,m_j\equiv n_j\equiv=\nu=2,\,I=J=5000,\, j(i)\equiv i,
$$
which results in $\dim x=20000$, $\dim y=45150$. The data matrices $q_i\succeq0,a_i,b_i,c_i$ were generated at random and normalized to have spectral norms 1, which ensures $\bV\leq1$. A generated instance was processed as follows:
\\
\indent $\bullet$ first, it was solved by the deterministic Mirror Prox algorithm (DMP) with on-line adjustable ``aggressive'' stepsize policy \cite{DMP}; up to this policy,
this is nothing but  SMP with $P_z$ specified as the unit mass sitting at $z$, $z\in Z$;
\\
\indent $\bullet$ next, it was solved by SMP (10 runs) with $k_x=1$, $k_y=100$ \footnote{with our $m,\nu,J$, the coefficient at $k_x$ in the right hand side of (\ref{costofG}) is nearly 30 times larger than the one at $k_y$, this is why we use $k_y\gg k_x$.} and the stepsize policy
$$
\gamma_\tau=\alpha\min\left[{1\over\sqrt{3}\cL}, {\sqrt{\Omega_X^2+\Omega_Y^2}\over\sqrt{7}\sigma\sqrt{\tau}}\right],\,\tau=1,2,...
$$
with $\cL$ and $\sigma$ given by (\ref{cLis}) (where we replace $\bV$ by its valid upper bound 1) and (\ref{sigmais}) (where we use $\Omega_X,\Omega_Y$ as given by (\ref{radii})). When $\alpha=1$, our stepsize policy becomes the ``rolling horizon'' version of  (\ref{stepsizes}); it can be shown that this policy (which does not require the number $t$ of steps to be  chosen in advance) is, theoretically, basically as good as its constant stepsizes prototype). The role of the ``acceleration factor'' $\alpha\geq1$ is to allow for larger stepsizes than those given by the worst-case-oriented considerations underlying (\ref{stepsizes}), the option which for DMP is given by the aforementioned on-line adjustable stepsize policy (in our experiments, the latter  resulted in stepsizes which, at average, were $\approx 250$ times the ``theoretically safe'' ones). The value of $\alpha$ we used (1000) was selected empirically in a small series of pilot experiments and was never revised in the main series of experiments.\\
\indent $\bullet$ In every experiment, a solution with the duality gap $\leq\epsilon=0.01$ was sought. Since the duality gap is not directly observable, this goal was achieved as follows. From time to time (specifically, after every 30 iterations for DMP and every 50 iterations for SMP) we computed  $F(z^t)$  for the current approximate solution $z^t=(x^t,y^t)$ (see (\ref{SMP})), thus getting $g:=\nabla_x\phi(x^t,y^t)$ and $\cA(x^t)=\nabla_y\phi(x^t,y^t)$. We then computed the maximal eigenvalue $\phi^+=\lambda_{\max}(\cA(x^t))$, which is nothing but $\overline{\phi}(x^t)=\max_{y\in Y}\phi(x,y)$, and the quantity  $\phi^-=\min_{x\in X}[\phi(x^t,y^t)+\Tr([x-x^t]^Tg)]$, which is a lower bound on $\underline{\phi}(y^t)=\min_{x\in X}\phi(x,y^t)$. The quantity $\Delta=\phi^+-\phi^-$ is an upper bound on $\DualityGap(x^t,y^t)$, and the relation $\Delta\leq\epsilon=0.01$ was used as the termination criterion.
\par
The results of a typical experiment are presented in table \ref{table1}. We see that while randomization increases essentially the iteration count, it results in overall reduction of the CPU time by a quite significant factor. It makes sense to note that of 2167 sec CPU time for DMP, 91\% (1982 sec) were spent on computing the values  of $F$, and just 9\% -- on computing prox-mappings; for SMP, both these components take nearly equal times.
\begin{table}
\centerline{
\begin{tabular}{|c||c|c|c||c|c|c||}
\cline{2-7}
\multicolumn{1}{c||}{}&\multicolumn{3}{|c||}{Iteration count}&\multicolumn{3}{|c||}{CPU, sec}\\
\hline
Algorithm&$\min$&mean&$\max$&$\min$&mean&$\max$\\
\hline\hline
DMP&\multicolumn{3}{c||}{61}&\multicolumn{3}{c||}{2167}\\
\hline
SMP&251&281&351&496&571&708\\
\hline\hline
\end{tabular}}
\caption{\label{table1}. Effect of randomization, problem (\ref{SPI}) ($I=J=5000,m=300,j(i)\equiv i,m_j\equiv n_j\equiv 2$). In the table: DMP/SMP -- Deterministic/Randomized
Mirror Prox. Data for SMP are obtained in 10 runs of the algorithm. Running times include those needed to check the termination criterion.}
\end{table}
\subsection{Illustration II: low dimensional approximation}
Consider the problem as follows: we are given $n$ {\sl unit} vectors $a_j\in\bR^m$, $1\leq j\leq n$, and know that for some given $k$, $1< k\leq m/2$, and $\delta\in(0,1)$ all $a_j$'s are at the $\|\cdot\|_2$-distance at most $\delta<1$ form certain $k$-dimensional subspace $L$, common for all points. The problem is to recover this subspace\footnote{Note the difference with the PCA -- Principal Component Analysis: we want to minimize the maximal, over $j$, deviation of $a_j$, from $L$ rather than the sum of squares of these deviations.},  which reduces to solving the problem
\begin{equation}\label{intract}
\Opt_*=\max_{x\in \cP_k}\min_{y\in Y} \sum_{j=1}^n y_ja_j^Txa_j,
\end{equation}
where $\cP_k\subset E_x=\bS^m$ is the family of all orthoprojectors of rank $k$ on $\bR^p$, and $Y=\{y\in\bR^n_+:\sum_jy_j=1\}$ is the standard simplex in $E_y=\bR^m$. The set $\cP_k$ is nonconvex; we relax it to the set
$$
X=\{x\in\bS^m: I_m\succeq x\succeq0,\Tr(x)=k\},
$$
thus arriving at the relaxed saddle point problem
\begin{equation}\label{(SP)}
\begin{array}{c}
-\Opt=\min_{x\in X}\max_{y\in Y}[\phi(x,y):=-\sum_{j=1}^n y_ja_j^Txa_j]\\
F_x(x,y)=-\sum_{j=1}^n y_ja_ja_j^T,\quad F_y(x,y)=[a_1^Txa_1;...;a_n^Txa_n]\\
\end{array}
\end{equation}
(we have equivalently transformed the relaxed problem to fit our standard notation). Note that $\phi$ is a polynomial of degree $d=2$ (just bilinear). Let us apply to (\ref{(SP)}) our approach.
\paragraph{Scale factor.} We clearly have $\bV\leq1$ (recall that $\|a_j\|_2=1$, $0\preceq x\preceq I_m$ for $x\in X$, and $\|y\|_1\leq1$ for $y\in Y$).
\paragraph{Setup.} We set
\begin{equation}\label{omegasLDA}
\begin{array}{rcl}
\omega_X(x)&=&{8\over q(1+q)}\sum_{i=1}^m\lambda_i^{1+q}(x),\,\,q=\min[1,\ln(k)/\ln(m/k)],\\
\omega_Y(y)&=&{8\sqrt{\e}\over p(1+p)}\sum_{j=1}^n y_j^{1+p},\,p=1/(2\ln(n)),\\
\end{array}
\end{equation}
thus getting d.-g.f.'s for $X$, $Y$ compatible with $\|\cdot\|_X$, $\|\cdot\|_Y$, respectively (Proposition \ref{propNN1} and Remark \ref{remlast}), the corresponding radii of $X$, $Y$ are
\begin{equation}\label{radiiNew}
\Omega_x\leq O(1)\sqrt{{k\ln(k)/\ln(m/k)}},\,\,\Omega_Y\leq O(1)\sqrt{\ln(n)},
\end{equation}
see (\ref{XomegaX}).
\paragraph{Deterministic algorithm.} When solving (\ref{(SP)}) within accuracy $\epsilon<1$ by the deterministic algorithm DMP,\\
--- the iteration count is $N_\det(\epsilon)=O(1) {k\ln(k)/\ln(m/k)+\ln(n)\over\epsilon}$,\\
--- the complexity of an iteration is $O(1)(m^3+n)$ a.o. for computing prox-mappings and $O(1)m^2n$ a.o. for computing the values of $F$.
\\
Note that {\sl as far as deterministic solution algorithms are concerned}, the outlined bounds result in the best known to us overall arithmetic complexity of finding an $\epsilon$-solution in the large scale case.
\par
When $n\gg m$, the cost of prox-mapping is much smaller than the one of computing the values of $F$, implying that there might be room for accelerating by randomization.
\paragraph{Randomization.} In order to compute, given $z=(x,y)\in X\times Y$, unbiased random estimates of $F_x(x,y)$ and $F_y(x,y)$, we act as follows.
\begin{enumerate}
\item We associate with $y$ the distribution $P_y$ on $Y$ as follows: $\eta\sim P_y$ takes the values $e^j$ (basic orths in $\bR^n$) with probabilities $y_j$, $1\leq j\leq n$ (cf. Example 1); the corresponding random estimate $G^x$ of $F_x(x,y)$ takes the values $-a_ja_j^T$ with probabilities $y_j$, $1\leq j\leq n$. Generating the estimate requires  the ``setup cost'' of $O(n)$ a.o.; after this cost is paid, generating  the estimate takes $O(1)[\ln(n)+m^2]$ a.o.
\item We associate with $x\in X$ the distribution $P_x$ on $X$ as follows. Given $x$, we compute its eigenvalue decomposition $x=U\Diag\{\xi\}U^T$. The vector $\xi$ belongs to the polytope $Q=\{\xi\in\bR^m:0 \leq \xi_i\leq 1\,\forall i, \sum_i\xi_i=k\}$. Now, there is a simple algorithm \cite[section A.1]{Rand} which allows, given  $\xi\in Q$, to represent $\xi$ as a convex combination $\sum_{i=1}^m\lambda_i\xi^i$ of extreme points of $Q$ (which are Boolean vectors with exactly $k$ entries equal to 1); the cost of building this representation is $O(1)km^2$ a.o. We build this representation and define $P_x$ as the distribution of a random symmetric matrix which takes values  $U\Diag\{\xi^i\}U^T$ with probabilities $\lambda_i$, $1\leq i\leq m$, so that the random estimate of $F_y(x,y)$ is the vector with the entries $G^y_j=\sum_{\ell\in I_i}(a_j^T\Col_\ell[U])^2$, $1\leq j\leq n$, where $I_i$ is the set of indexes of the $k$ nonzero entries of the Boolean vector $\xi^i$, and $i$ takes values $1,...,m$ with probabilities $\lambda_1,...,\lambda_m$. Finally, we set $P_z=P_x\times P_y$. Note that this distribution is supported on $X\times Y$ (i.e., Assumption B is satisfied with $\rho=0$). The ``setup'' cost of sampling from $P_x$ is $O(1)m^3$ a.o.; after this cost is paid, generating a sample value of $G^y$ costs $O(1)kmn$ a.o.
\end{enumerate}
With the outlined randomization, the cost of generating a sample value of $G_{k_x,k_y}$  in the range $\ln(n)\leq O(1)m^2$ costs
$$
O(1)(m^3+k_xkmn+k_ym^2) \hbox{\ \rm a.o.}
$$
When $n\gg m\gg k$ and $k_x$, $k_y$ are moderate, this cost is by far less than the cost $O(1)m^2n$ of deterministic computation of $F(x,y)$, so that our randomization indeed possesses some potential. Analysis completely similar to the one in section  \ref{Ill1} shows that our current situation is completely similar to the one in the latter section: while with $k_x=O(1)$, $k_y=O(1)$, the iteration count for the randomized algorithm is proportional to $\epsilon^{-2}$ instead of being proportional to $\epsilon^{-1}$, as for the deterministic algorithm, the growth in this count, in certain meaningful range of values of $k,m,n,\epsilon$ is by far overweight by reduction in the cost of an iteration. As a result, for $\epsilon$ fixed and in the case of appropriate proportion between $k,m,n$, the randomized algorithm progressively outperforms its deterministic competitor as the sizes of the problem grow.
\paragraph{Numerical illustration.}
In the experiment we are about to describe, the sizes of problem (\ref{(SP)}) were selected as
$$
m=100,\,k=10,\,n=300,000.
$$
The data points $a_j$ were selected at random in certain ``smart'' way aimed at creating difficult instances; we are not sure that this goal was indeed achieved, but at least the PCA solution (which, with straightforward random generation of $a_j$, turns out to recover perfectly well the approximating subspace) was ``cut off:'' -- the largest, over all $j$, distance of $a_j$'s to the $k=10$-dimensional PCA subspace in our experiments was as large as 0.99. \par
Implementation of the approach was completely similar to the one outlined in section \ref{Ill1}; the only specific issue which should be addressed here is the one of termination. Problem (\ref{(SP)}) by its origin is no more than a relaxation of the ``true'' problem (\ref{intract}), so solving it within a given accuracy is of no much interest. Instead, we from time to time (namely, every 10 iterations) took the $x$-component $x^t$ of the current approximate solution, subject it to eigenvalue decomposition and checked straightforwardly what is the largest, over $j\leq n$, $\|\cdot\|_2$-deviation $D$ of $a_j$ from the $k$-dimensional subspace of $\bR^m$ spanned by $k$ principal eigenvectors of $x^t$. We terminated the solution process when this distance was $\leq \delta+\epsilon$, where $\epsilon$ is a prescribed tolerance.
\par
Typical experimental results are presented in table \ref{table2}. The results look surprisingly good -- the iteration count is quite low and is the same for both deterministic and randomized algorithms. We do not know whether this unexpected phenomenon reflects the intrinsic simplicity of the problem,  or our inability to generate really difficult instances, or the fact that we worked with although reasonable, but not ``really small'' values of $\epsilon$; this being said, we again see that randomization reduces the CPU time by a quite significant factor.
\begin{table}
\centerline{\small\begin{tabular}{|c|c|c|c|c|}
\cline{2-5}
\multicolumn{1}{c|}{}&Method&\# of steps&CPU, sec& Final deviation $D$\\
\hline\hline
$\delta=0.4,\delta+\epsilon=0.45$&DMP&20&{478}&0.401\\
\hline
\multicolumn{1}{c|}{}&SMP&20&{104}&0.427\\
\hline\hline
$\delta=0.6,\delta+\epsilon=0.65$&DMP&20&{ 504}&0.603\\
\hline
\multicolumn{1}{c|}{}&SMP&20&{ 105}&0.620\\
\hline\hline
$\delta=0.8,\delta+\epsilon=0.85$&DMP&20&{478}&0.809\\
\hline
\multicolumn{1}{c|}{}&SMP&20&{92}&0.819\\
\cline{2-5}
\end{tabular}}
\caption{\label{table2} Deterministic (DMP) and randomized (SMP, $k_x=1$, $k_y=10$) algorithms on the low dimensional approximation problem. }
\end{table}

\appendix
\section{Proofs}
\subsection{Proof of Lemma \ref{lem1}}
In what follows, $C_i$ are positive quantities depending solely on $d$, and $\widehat{Z}$ is the convex hull of $\{0\}\cup Z$.
Observe that $L[\widehat{Z}]:=\Lin(\widehat{Z}-\widehat{Z})\supset L[Z]:=\Lin(Z-Z)$ and $\widehat{Z}^s:={1\over 2}[\widehat{Z}-\widehat{Z}]\supset Z^s:={1\over 2}[Z-Z]$; as a result,
\begin{equation}\label{emma}
\|z\|_{\widehat{Z}}\leq\|z\|\,\,\forall z\in L[Z].
\end{equation}
\paragraph{1$^0$.}  Observe that for some $C_1$ one has
\begin{equation}\label{C1}
\forall (z\in \widehat{Z},2\leq k\leq d): |Q_k(z,...,z)|\leq C_1\bV.
\end{equation}
Indeed, let $z\in \widehat{Z}$. The univariate polynomial
$$
p(t):=\widehat{\phi}(tz)=\sum_{k=2}^d Q_k(z,...,z)t^k
$$
on the segment $0\leq t\leq 1$ is bounded in absolute value by $\bV$ (since $\bV$ is the variation of $\widehat{\phi}$ on $\widehat{Z}\ni0$ and $\widehat{\phi}(0)=0$), so that the moduli  $|Q_k(z,...,z)|$ of its coefficients are bounded by $C_1\bV$ for some $C_1$ depending solely on $d$.
\paragraph{2$^0$.} Our next observation is that for some $C_2$ one has
\begin{equation}\label{C2}
\forall (z\in L[\widehat{Z}], 2\leq k\leq d): |Q_k(z,...,z)|\leq C_2\bV\|z\|_{\widehat{Z}}^k.
\end{equation}
Indeed, let $2\leq k\leq d$. By homogeneity it suffices to verify (\ref{C2}) when $\|z\|_{\widehat{Z}}=1$, so that $z={1\over 2}[z^1-z^2]$ with some $z^1,z^2\in \widehat{Z}$. Setting $h(t_1,t_2)=t_1z^1+t_2z^2$, consider the polynomial of two variables
$$
p(t_1,t_2)=Q_k(h(t_1,t_2),h(t_1,t_2),...,h(t_1,t_2)).
$$
$p$ is a polynomial of degree $\leq k\leq d$ on the 2D plane which is bounded in absolute value by $C_1\bV$ in the triangle $t_1,t_2\geq 0, t_1+t_2\leq1$ (by (\ref{C1}) combined with the  fact that for the outlined $t_1,t_2$ we have $h(t_1,t_2)=(1-t_1-t_2)\cdot 0+t_1z^1+t_2z^2\in \widehat{Z}$). As a result, the moduli of the coefficients of $p$ do not exceed $C_3\bV$ with appropriately chosen $C_3$, whence $p(1/2,-1/2)=Q_k(z,...,z)$ is bounded in absolute value by $C_2\bV$ with appropriately chosen $C_2$.
\paragraph{3$^0$.} Now let $2\leq k\leq d$, and let $z^1,...,z^k\in L[\widehat{Z}]$, $\|z^i\|_{\widehat{Z}}\leq1$, $1\leq i\leq k$. Consider the polynomial of $k$ real variables
$$
p(t_1,...,t_k)=Q_k(\sum_{i=1}^kt_iz^i,\sum_{i=1}^kt_iz^i,...,\sum_{i=1}^kt_iz^i).
$$
The degree of this polynomial does not exceed $k\leq d$, and
$$
|p(t_1,...,t_k)|\leq C_2\bV\|t_1z^1+...+t_kz^k\|_{\widehat{Z}}^k\leq C_2\bV\|t\|_1^k
$$
by (\ref{C2}).
It follows that for some $C_4$ we have
$$
\left|{\partial^k p(t_1,...,t_k)\over\partial t_k\partial t_{k-1}...\partial t_1}\right|\leq C_4\bV
$$
The left hand side in this relation is $k!|Q_k(z^1,...,z^k)|$ (recall that $Q_k(\cdot,...,\cdot)$ is $k$-linear and symmetric),
and we see that
$$
\forall \{z^i\in L[\widehat{Z}],\|z^i\|_{\widehat{Z}}\leq1\}_{i=1}^k: |Q_k(z^1,...,z^k)|\leq {C_4\over k!}\bV,
$$
which by homogeneity implies (\ref{eq1}).
\paragraph{4$^0$.} It remains to prove the ``in particular'' part of Lemma \ref{lem1}. Taking into account (\ref{unitary}), (\ref{direct1}) --- (\ref{direct3}), to this end it suffices to verify that
the second order directional derivative $D^2\phi(z)[h,h]={d^2\over dt^2}\big|_{t=0}\phi(z+th)$ taken at a point $z\in Z$ along a direction $h\in L[Z]$ satisfies
$$
|D^2\phi(z)[h,h]|\leq \cL\|h\|^2
$$
with $\cL$ given by (\ref{cLis}). This is immediate: by (\ref{phinew}) we have
$$
D^2\phi(z)[h,h]=\sum_{k=2}^dk(k-1)Q_k(h,h,z,...,z).
$$
We have $\|z\|_{\widehat{Z}}\leq2$ by definition of $\|\cdot\|_{\widehat{Z}}$ (recall that $z\in Z$), so that by (\ref{eq1}) the modulus of the right hand side does not exceed $\sum_{k=2}^d k(k-1)2^{k-2}\|h\|_{\widehat{Z}}^2C^{(1)}\bV$. It remains to note that $\|h\|_{\widehat{Z}}\leq\|h\|$ due to $h\in L[Z]$ and (\ref{emma}). \qed
\subsection{Proof of Lemma \ref{lemraash}}
\paragraph{1$^0$.} Let, as always, $\widehat{Z}$ be the convex hull  of $\{0\}\cup Z$, and let us fix $z\in Z$. Consider the random vectors $\zeta_x,\zeta_y$ taking values in $E_x$, $E_y$, respectively:
$$
\begin{array}{l}
\zeta_x=G_x[z^1,...,z^{d-1}],\,\zeta_y=G_y[z^1,...,z^{d-1}],\,\zeta=[\zeta_x;\zeta_y],\\
\delta_x=\zeta_x-F_x(z),\,\delta_y=\zeta_y-F_y(z),\,\delta=[\delta_x;\delta_y],\\
\end{array}
$$
where $z^1,...,z^{d-1}$ are drawn, independently of each other, from $P_z$. We claim that for some $C_5$, depending solely on $d$, it holds
\begin{equation}\label{step1}
\|\delta\|_*\leq C_5\bV(1+\rho)^{d-1}.
\end{equation}
Indeed, by construction of $G[z^1,...,z^{d-1}]$ and in view of (\ref{phinew}) we have
\begin{equation}\label{chain1}
\begin{array}{ll}
&\forall h\in L[Z]: \left\{\begin{array}{rcl}\langle \zeta,h\rangle&=&\sum_{k=1}^dk Q_k(Dh,z^1,...,z^{k-1})\\
\langle F(z),h\rangle& =&\sum_{k=1}^dk Q_k(Dh,z,...,z)\\
\end{array}\right.\\
\Rightarrow&\|\delta\|_*=\max\limits_{h\in L[Z],\|h\|\leq1}\sum_{k=1}^dk[Q_k(Dh,z,...,z)
-Q_k(Dh,z^1,...,z^{k-1})]\\
&\leq\max\limits_{h:\|h\|\leq1}\sum_{k=2}^dk C^{(1)}\bV[\|Dh\|_{\widehat{Z}}[\|z\|_{\widehat{Z}}^{k-1}+\|z^1\|_{\widehat{Z}}\|z^2\|_{\widehat{Z}}...
\|z^{k-1}\|_{\widehat{Z}}],\\
\end{array}
\end{equation}
where the concluding inequality is due to (\ref{eq1}) (take into account that $h\in L[Z]=L[X]\times L[Y]$, whence $Dh\in L[Z]\subset L[\widehat{Z}]$, and that $z,z^i\in\Aff(Z)\subset L[\widehat{Z}]$). Invoking (\ref{emma}), we get  $\|Dh\|_{\widehat{Z}}\leq \|Dh\|=\|h\|$. Besides this,  $z\in Z$ implies that $\|z\|_{\widehat{Z}}\leq2$, while Assumption B combines with (\ref{emma})
and the relation $\|z'\|_{\widehat{Z}}\leq 2$ for all $z'\in Z$ to imply that $\|z^i\|_{\widehat{Z}}\leq 2(1+\rho)$.  In view of these observations, the concluding quantity in (\ref{chain1}) is $\leq \sum_{k=2}^d kC^{(1)}\bV2^{k-1}[1+(1+\rho)^{k-1}]$, so that
 $\|\delta\|_*\leq C_5\bV(1+\rho)^{d-1}$ with $C_5$ depending solely on $d$, as claimed in (\ref{step1}).
\paragraph{2$^0$.} We need the following fact:
\begin{proposition}\label{martin} Let $F$ be a Euclidean space, $\|\cdot\|$ be a norm on $F$, $\|\cdot\|_*$ be the conjugate norm, let $\Xi$ be a Polish space  equipped with a Borel probability distribution, and $\cF$ be the space of all Borel mappings $f:\Xi\to F$ such that for some $c_f\in(0,\infty)$ it holds $\bE\{\exp\{\|f(\cdot)\|_*^2/c_f^2\}\}\leq\exp\{1\}$. Then
\par{\rm (i)} $\cF$ is a linear space, and the quantity $\sigma[f]=\inf\{c>0: \bE\{\exp\{\|f(\cdot)\|_*^2/c^2\}\}\leq\exp\{1\}\}$ is a (semi)norm on $\cF$;
\par
{\rm (ii)} Let $U$ be a convex compact set in $F$ such that $U^s={1\over2}[U-U]$ is the unit ball of the norm $\|\cdot\|$. Assume that $U$ admits a d.-g.f. $\omega(\cdot)$ compatible with $\|\cdot\|$, and let $\Omega$ be the $\omega$-radius of $U^s$. Then for properly chosen absolute constant $O(1)$, with $\chi=O(1)\Omega$ the following holds true:
\begin{quote}
(!) Let $f_1,f_2,...$ be an $F$-valued martingale-difference, that is, a sequence of random vectors taking values in $F$ and such that $\bE_{|t-1}\{f_t\}\equiv0$ for all $t$, where $\bE_{|t-1}$ is the conditional expectation w.r.t. the $\sigma$-algebra spanned by $f_1,...,f_{t-1}$.  Assume that for a sequence of nonnegative deterministic reals $\sigma_1,\sigma_2,...$ it holds
$$
\bE_{|t-1}\{\exp\{f_t(\cdot)\|_*^2/\sigma_t^2\}\}\leq \exp\{1\} \hbox{\ a.s.}
$$
Then for every $t$ one has
\begin{equation}\label{sigmamart}
\sigma[f_1+...+f_t]\leq  \chi\sqrt{{\sum}_{\tau=1}^t\sigma_\tau^2}.
\end{equation}
\end{quote}
\end{proposition}
{\bf Proof.} (i) is well known; for the sake of completeness, here is the proof. The fact which indeed needs verification is the triangle inequality. Thus, let $f,g\in\cF$, $a>\sigma[f]$ and $b>\sigma[g]$; all we need is to prove that $a+b\geq\sigma[f+g]$.  Setting $\lambda=a/(a+b)$, we have
$$\begin{array}{l}
\exp\{\|f+g\|_*^2/(a+b)^2\}\leq\exp\{[\|f\|_*+\|g\|_*]^2/(a+b)^2\}\\
=\exp\{[\lambda(\|f\|_*/a)+(1-\lambda)(\|g\|_*/b)]^2\}
\leq \lambda\exp\{(\|f\|_*/a)^2\}+(1-\lambda)\exp\{(\|g\|_*/b)^2\},\\
\end{array}
$$  where the concluding $\leq$ is due to the convexity of the univariate function $\exp\{s^2\}$. Taking expectations in the resulting inequality, we get
$
\bE\{\exp\{\|f+g\|_*^2/(a+b)^2\}\leq \exp\{1\}$, that is, $a+b\geq\sigma[f+g]$, as claimed. (i) is justified.\par
 (ii): Let $\psi(u)=\left\{\begin{array}{ll}\omega(2u),u\in {1\over 2}U\\
 +\infty,u\not\in {1\over 2}U\\
 \end{array}\right.$, so that $\Dom \psi={1\over 2}U$, and let  $f(\xi)=\max_{u\in {1\over 2}U}[\langle \xi,u\rangle -\psi(u)]$ be the Fenchel transform of $\psi$. Since $\omega$ is strongly convex on $U$, modulus 1, w.r.t. $\|\cdot\|$, $\psi$ is strongly convex on its domain, modulus $4$ w.r.t. $\|\cdot\|$, whence, by the standard properties of the Fenchel transformation, $f$ possesses Lipschitz continuous gradient, specifically, $\|f'(\xi)-f'(\eta)\|\leq \|\xi-\eta\|_*/4$  for all $\xi,\eta$. The Fenchel transform of the function $\psi_-(u)=\psi(-u)$ is $f_-(\xi)=f(-\xi)$. Now let $\psi^s$ be the $\inf$-convolution of $\psi(\cdot)$ and $\psi_-(\cdot)$, i.e., the function
 $$\begin{array}{rcl}
 \psi^s(u)&=&\inf_{v,w:v+w=u}\left(\psi(v)+\psi_-(w)\right)=\inf_{v,w':v-w'=u}(\psi(v)+\psi(w'))\\
 &=&
 \left\{\begin{array}{ll} \min_{v,w'\in{1\over 2}U:v-w'=u}[\psi(v)+\psi(v')],&u\in U^s={1\over 2}[U-U]\\
 +\infty,&u\not\in U^s\\
 \end{array}\right.\\
 \end{array}
 $$
 The Fenchel transform of the $\inf$-convolution of $\psi$ and $\psi_-$ is the sum of the Fenchel transforms of $\psi$ and $\psi_-$ (recall that the functions are convex with closed compact domains and are continuous on their domains), that is, it is the function $g(\xi)=f(\xi)+f(-\xi)$, so that the Fenchel transform of $\psi^s(\cdot)$ satisfies $\|g'(\xi)-g'(\eta)\|\leq \|\xi-\eta\|_*/2$. By the standard properties of the Fenchel transform, it follows that $\psi^s(\cdot)$ is strongly convex, modulus $2$ w.r.t. $\|\cdot\|$, on its domain (which is exactly the unit ball $U^s$ of the norm $\|\cdot\|$), and the variation (the maximum minus the minimum) of $\psi^s$ on the domain is $\leq\Omega^2$ (since the variation of $\psi(\cdot)$ over ${1\over 2}U$, that is, the variation of $\omega(\cdot)$ over $U$, is $\Omega^2/2$).
 The bottom line is that the unit ball $U^s$ of $\|\cdot\|$ admits a continuous strongly convex, modulus 1 w.r.t. $\|\cdot\|$, function (specifically, ${1\over2}\psi^s\big|_{U^s}$) with variation over $U^s$ not exceeding $\Omega^2/2$. Invoking \cite[Proposition 3.3]{LarDev}, it follows that the space $(F,\|\cdot\|_*)$ is $O(1)\Omega^2$ regular (for details, see \cite{LarDev}). With this in mind, the conclusion (!) in (ii) is an immediate consequence of \cite[Theorem 2.1.(ii)]{LarDev}.     \qed
\paragraph{3$^0$.} Now we can complete the proof of Lemma \ref{lemraash}. We have already seen that \SO generates unbiased random estimates of $F$, whence $\SOmath_{k_x,k_y}$ possesses the same property; thus, $\SOmath_{k_x,k_y}$ meets the requirement (\ref{aresuch}.$b$), which is the first claim in Lemma \ref{lemraash}. Now let us prove the second claim in this Lemma. In the notation from item 1$^0$, setting $F=L[X]$ and denoting by $\pi$ the orthoprojector of $E_x$ onto $F\subset E_x$, (\ref{step1}) implies that
\begin{equation}\label{medium}
\|\pi\delta_x\|_{X,*}=\|\delta_x\|_{X,*}\leq C_5\bV(1+\rho)^{d-1}
 \end{equation}
 (since $\|\delta\|_*=\|\delta_x\|_{X,*}+\|\delta_y\|_{Y,*}$). The $x$-component $\Delta_x$ of the ``observation error'' of $\SOmath_{k_x,k_y}$ (the difference $\Delta=[\Delta_x;\Delta_y]$ of the random estimate of $F(z)$ generated by $\SOmath_{k_x,k_y}$ and  $F(z)$) is
\begin{equation}\label{fs}
\Delta_x=\sum_{t=1}^{k_x}f_t\Rightarrow \pi\Delta_x=\sum_{t=1}^{k_x}\widetilde{f}_t,
\end{equation}
where $\widetilde{f}_1,...,\widetilde{f}_{k_x}$ are independent copies of the zero mean random vector $k_x^{-1}\pi \delta_x\in F$. Besides this, choosing a point $\bar{x}\in X$ and setting $\widetilde{X}=X-\bar{x}\subset F$, $\widetilde{\omega}(\xi)=\omega(\bar{x}+\xi)$, $\xi\in \widetilde{X}$, we see that $X^s={1\over 2}[\widetilde{X}-\widetilde{X}]$ admits a d.-g-f., specifically, $\widetilde{\omega}(\cdot)$, which is compatible with $\|\cdot\|_X$ and is such that the $\widetilde{\omega}$-radius of $\widetilde{X}$ is $\Omega_X$. Invoking Proposition \ref{martin}.(ii) and taking into account that we are in the situation $\sigma[\widetilde{f}_t]=\sigma[f_t]\leq C_5k_x^{-1}\bV(1+\rho)^{d-1}$ by (\ref{medium}), we get that for properly chosen $C_6$ depending solely on $d$ we have
$$
\bE\left\{\exp\{\|\Delta_x\|_{X,*}^2/\widetilde{\sigma}_x^2\}\right\}\leq\exp\{1\},\,\,\widetilde{\sigma}_x=C_6\Omega_X\bV(1+\rho)^{d-1}/\sqrt{k_x}
$$
(note that $\|\Delta_x\|_{X,*}=\|\pi\Delta_x\|_{X,*}$).
Besides this, by (\ref{medium}) $\|\widetilde{f}_t\|_{X,*}\leq C_5k_x^{-1}\bV(1+\rho)^{d-1}$ almost surely, whence
$$
\bE\left\{\exp\{\|\Delta_x\|_{X,*}^2/\bar{\sigma}_x^2\}\right\}\leq\exp\{1\},\,\,\bar{\sigma}_x=C_5\bV(1+\rho)^{d-1}.
$$
The bottom line is that with properly selected $C_7$ depending solely on $d$ and with
$$\sigma_x=C_7\bV(1+\rho)^{d-1}\min[1,\Omega_X/\sqrt{k_x}]$$
we have
$$
\bE\left\{\exp\{\|\Delta_x\|_{X,*}^2/\sigma_x^2\}\right\}\leq\exp\{1\}.
$$
By similar reasons,  with properly selected $C_8$ depending solely on $d$ and with
$$\sigma_y=C_8\bV(1+\rho)^{d-1}\min[1,\Omega_Y/\sqrt{k_y}]$$
we have
$$
\bE\left\{\exp\{\|\Delta_y\|_{Y,*}^2/\sigma_y^2\}\right\}\leq\exp\{1\}.
$$
Taking into account that $\|\Delta\|_*=\|\Delta_x\|_{X,*}+\|\Delta_y\|_{Y,*}$ and item (i) of Proposition \ref{martin}, the second claim in Lemma \ref{lemraash} follows. \qed
\subsection{Proofs for section \ref{sectIll}}
What follows is a slight modification of the reasoning from \cite[Section A.2]{NesNem2013}; we present it here to make the paper self-contained.
\paragraph{A.} Let $\bS^m$ be the space of $m\times m$ symmetric matrices equipped with the Frobenius inner product; for $y\in \bS^m$, let $\lambda(y)$ be the vector of eigenvalues of $y$ (taken with their multiplicities in the non-ascending order). For an integer $k$, $1\leq k\leq m$, let $Y^k=\{y\in \bS^m: \|\lambda(y)\|_\infty\leq1,\|\lambda(y)\|_1\leq k\}$, so that $Y^k$ is the unit ball of certain rotation-invariant norm $\|\cdot\|_{(k)}$ on $\bS^m$.
\begin{lemma}\label{LemNN} Let  $m,n,k$ be integers such that $m\geq n\geq k\geq1$, and let $F$ be a linear subspace in $\bS^m$ such that every matrix $y\in F$ has at most $n$ nonzero eigenvalues. Let, further, $q\in(0,1)$, and let
$$
\chi(y)={1\over 1+q}\sum_{j=1}^m|\lambda_j(y)|^{1+q}:\bS^m\to\bR.
$$
The function $\chi(\cdot)$ is continuously differentiable, convex, and its restriction on the set $Y_F^k=\{y\in F:\|y\|_{(k)}\leq1\}$ is strongly convex w.r.t.  $\|\cdot\|_{(k)}$ modulus
\begin{equation}\label{beta}
\beta=q\min[1,{1\over 2}k^{1+q}n^{-q}].
\end{equation}
\end{lemma}
{\bf Proof.} {\bf1$^0$.}
Observe that
\begin{equation}\label{chiofy}
\chi(y)=\Tr(f(y)),\,f(s)={1\over 1+q}|s|^{1+q}.
\end{equation}
Function $f(s)$ is continuously differentiable on the axis and twice continuously differentiable outside of the origin; consequently, we can find a sequence of polynomials $f_r(s)$ converging, as $r\to\infty$,  to $f$ along with their first derivatives uniformly on every compact subset of $\bR$ and, besides this, converging to $f$ uniformly along with the first and the second derivative on every compact subset of $\bR\backslash\{0\}$. Now let $y,h\in \bS^m$, let $y=u\Diag\{\lambda\} u^T$ be the eigenvalue decomposition of $y$, and let $h=u\widehat{h}u^T$.  For a polynomial $p(s)=\sum_{\ell=0}^Lp_\ell s^\ell$, setting $P(w)=\Tr(\sum_{\ell=0}^Lp_\ell w^\ell):\bS^m\to\bR$, and denoting by $
\gamma$ a closed contour in $\bC$ encircling the spectrum of $y$, we have
$$
\begin{array}{ll}
(a)&P(y)=\Tr(p(y))=\sum_{j=1}^mp(\lambda_j(y))\\
(b)&DP(y)[h]=\Tr(\sum_{\ell=0}^L\ell p_\ell\Tr(y^{\ell-1}h))=\Tr(p'(y)h)=\sum_{j=1}^mp'(\lambda_j(y))\widehat{h}_{jj}\\
(c)&D^2P(y)[h,h]={d\over dt}\big|_{t=0}DP(y+th)[h]={d\over dt}\big|_{t=0}\Tr(p'(y+th)h)\\
&={d\over dt}\big|_{t=0}{1\over 2\pi\imath}\oint\limits_\gamma \Tr(h(zI-(y+th))^{-1})p'(z)dz={1\over 2\pi\imath}\oint\limits_\gamma\Tr(h(zI-y)^{-1}h(zI-y)^{-1})p'(z)dz\\
&={1\over 2\pi\imath}\oint\limits_\gamma \sum_{i,j=1}^m\widehat{h}_{ij}^2{p'(z)\over(z-\lambda_i(y))(z-\lambda_j(y))}dz=\sum_{i,j=1}^n\widehat{h}_{ij}^2\Gamma_{ij},\\
&\Gamma_{ij}=\left\{\begin{array}{ll} {p'(\lambda_i(y))-p'(\lambda_j(y))\over\lambda_i(y)-\lambda_j(y)},&\lambda_i(y)\neq\lambda_j(y)\\
p''(\lambda_i(y)),&\lambda_i(y)=\lambda_j(y)\\
\end{array}\right.
\end{array}
$$
We conclude from $(a,b)$ that as $k\to\infty$, the real-valued polynomials $F_r(\cdot)=\Tr(f_r(\cdot))$ on $\bS^m$ converge, along with their first order derivatives,
uniformly on every bounded subset of $\bS^m$, and the limit of the sequence, by $(a)$, is exactly $\chi(\cdot)$. Thus, $\chi(\cdot)$ is continuously differentiable, and $(b)$ says that
\begin{equation}\label{frstder}
D\chi(y)[h]=\sum_{j=1}^mf'(\lambda_j(y))\widehat{h}_{jj}.
\end{equation}
Besides this, $(a$-$c)$ say that if $U$ is a closed convex set in $\bS^m$ which does not contain singular matrices, then $F_r(\cdot)$, as $r\to\infty$, converge  along with the first and the second derivative uniformly on every compact subset of $U$, so that $\chi(\cdot)$ is twice continuously differentiable on $U$, and at every point $y\in U$ we have
\begin{equation}\label{scndder}
D^2\chi(y)[h,h]=\sum_{i,j=1}^m\widehat{h}_{ij}^2\Gamma_{ij},\, \,\Gamma_{ij}=\left\{\begin{array}{ll} {f'(\lambda_i(y))-f'(\lambda_j(y))\over\lambda_i(y)-\lambda_j(y)},&\lambda_i(y)\neq\lambda_j(y)\\
f''(\lambda_i(y)),&\lambda_i(y)=\lambda_j(y)\\
\end{array}\right.
\end{equation}
and in particular $\chi(\cdot)$ is convex on $U$.
\par\noindent
{\bf 3$^0$.} We intend to prove that (i) $\chi(\cdot)$ is convex, and (ii) its restriction on the set $Y^k_F$ is strongly convex, with certain modulus $\alpha>0$, w.r.t. the norm $\|\cdot\|_{(k)}$. Since $\chi$ is continuously differentiable, all we need to prove (i) is to verify that
$$
\langle\chi'(y')-\chi'(y''),y'-y''\rangle\geq0\eqno{(*)}
$$
for a dense in $\bS^m\times\bS^m$ set of pairs $(y',y'')$, e.g., those with nonsingular $y'-y''$. For a pair of the latter type, the polynomial $q(t)=\Det(y'+t(y''-y'))$ of $t\in \bR$ is not identically zero and thus has finitely many roots on $[0,1]$. In other words, we can find finitely many points
$t_0=0<t_1<...<t_n=1$ such that all ``matrix intervals'' $\Delta_i=(y_i,y_{i+1})$, $y_k=y'+t_k(y''-y')$, $1\leq i\leq n-1$, are comprised of nonsingular matrices. Therefore $\chi$ is convex on every closed segment contained in one of $\Delta_i$'s, and since $\chi$ is continuously differentiable, $(*)$ follows.
\par\noindent
{\bf 4$^0$.}
It remains to prove that with $\beta$ given by (\ref{beta}) one has
\begin{equation}\label{target}
\langle\chi'(y')-\chi'(y''),y'-y''\rangle\geq\beta\|y'-y''\|_{(k)}^2\,\,\forall y',y''\in Y^k_F
\end{equation}
Let $\epsilon>0$, and let $Y^\epsilon$ be a convex open in $Y^k=\{y:\|y\|_{(k)}\leq 1\}$ neighborhood of $Y^k_F$ such that for all $y\in Y^\epsilon$ at most $n$ eigenvalues of $y$ are of magnitude  $>\epsilon$. We intend to prove that for some $\alpha_\epsilon>0$
one has
\begin{equation}\label{eq111}
\langle\chi'(y')-\chi'(y''),y'-y''\rangle\geq\alpha_\epsilon \|y'-y''\|_{(k)}^2\,\,\forall y',y''\in Y^\epsilon.
\end{equation}
Same as above, it suffices to verify this relation for a dense in $Y^\epsilon\times Y^\epsilon$ set of pairs $y',y''\in Y^\epsilon$, e.g., for those pairs $y',y''\in Y^\epsilon$ for which $y'-y''$ is nonsingular. Defining matrix intervals $\Delta_i$ as above and taking into account continuous differentiability of $\chi$, it suffices to verify that if $y\in\Delta_i$ and $h=y'-y''$, then
$D^2\chi(y)[h,h]\geq\alpha_\epsilon\|h\|_{(k)}^2$. To this end observe that by (\ref{scndder}) all we have to prove is that
\begin{equation}\label{pound}
D^2\chi(y)[h,h]=\sum_{i,j=1}^m\widehat{h}_{ij}^2\Gamma_{ij}\geq \alpha_\epsilon\|h\|_{(k)}^2.
\end{equation}
\par\noindent
{\bf 5$^0$.}
Setting $\lambda_j=\lambda_j(y)$, observe that $\lambda_i\neq 0$ for all $i$ due to the origin of $y$. We claim that if $|\lambda_i|\geq|\lambda_j|$, then $\Gamma_{ij}\geq q|\lambda_i|^{q-1}$. Indeed, the latter relation definitely holds true when $\lambda_i=\lambda_j$. Now, if $\lambda_i$ and $\lambda_j$ are of the same sign, then $\Gamma_{ij}={|\lambda_i|^q-|\lambda|_j^q\over|\lambda_i|-|\lambda_j|}\geq q|\lambda_i|^{q-1}$, since the derivative of the concave (recall that $0<q\leq1$) function $t^q$ of $t>0$ is positive and nonincreasing. If $\lambda_i$ and $\lambda_j$ are of different signs, then
$\Gamma_{ij}={|\lambda_i|^q+|\lambda_j|^q\over |\lambda_i|+|\lambda_j|}\geq |\lambda_i|^{q-1}$ due to $|\lambda_j|^q\geq|\lambda_j||\lambda_i|^{q-1}$, and therefore $\Gamma_{ij}\geq q|\lambda_i|^{q-1}$. Thus, our claim is justified.\par
W.l.o.g. we can assume that the positive reals $\mu_i=|\lambda_i|$, $i=1,...,m$, form a nondecreasing sequence, so that, by above, $\Gamma_{ij}\geq q\mu_j^{q-1}$ when $i\leq j$. Besides this, at most $n$ of $\mu_j$ are $\geq\epsilon$, since $y',y''\in Y^\epsilon$ and therefore $y\in Y^\epsilon$ by convexity of $Y^\epsilon$. By the above,
$$
D^2\chi(y)[h,h]\geq 2q\sum_{i<j\leq m}\widehat{h}_{ij}^2\mu_j^{q-1}+q\sum_{j=1}^m\widehat{h}_{jj}^2\mu_j^{q-1},
$$
or, equivalently by symmetry of $\widehat{h}$, if $$h^j=\left[\hbox{\scriptsize$\begin{array}{cccccc}&&&\widehat{h}_{1j}&&\\
&&&\widehat{h}_{2j}&&\\
&&&\vdots&&\\
\widehat{h}_{j1}&\widehat{h}_{j2}&\cdots&\widehat{h}_{jj}&&\\
&&&&&\\
&&&&&\\
\end{array}$}\right]$$
and $H_j$ is the Frobenius norm $\|h^j\|_\Fro$ of $h^j$, then
\begin{equation}\label{tthn}
D^2\chi(y)[h,h]\geq q\sum_{j=1}^mH_j^2\mu_j^{q-1}.
\end{equation}
\par\noindent
{\bf 6$^0$.}
Now note that
\begin{equation}\label{mujs}
0<\mu_j\leq 1\,\forall j,\,\mu_j\leq \epsilon,\, j\leq m-n,\,\,\sum_{j=1}^m\mu_j\leq k
\end{equation}
due to $y\in Y^\epsilon\subset Y^k$ and $\mu_j>0$ for all $j$. Now, by the definition of $\|\cdot\|_{(k)}$, setting
$$
\eta=\|h\|_{(k)}\,[\equiv\|\widehat{h}\|_{(k)}],
$$
observe that either $\eta$ is the spectral norm $\|\lambda(\widehat{h})\|_\infty$ of $\widehat{h}$, or $k\eta$ is the nuclear norm of $\widehat{h}$. In the first case, the Frobenius norm of $\widehat{h}$ is $\geq\eta$, meaning that $\sum_{j=1}^mH_j^2=\|\widehat{h}\|_\Fro^2\geq\eta^2$. Since $q\in(0,1)$ and $0<\mu_j\leq1$  for all $j$ by (\ref{mujs}), we conclude from (\ref{tthn}) and from the evident relation $\|\widehat{h}\|_\Fro^2=\sum_j\|h^j\|_\Fro^2=\sum_jH_j^2$ that in the case in question we have
 \begin{equation}\label{firstcase}
 D^2\chi(y)[h,h]\geq q\sum_{j=1}^mH_j^2\geq q\eta^2\equiv q\|h\|_{(k)}^2.
 \end{equation}
 Now assume that we are in the second case:
 \begin{equation}\label{secondcase_premise}
k \|h\|_{(k)}=k\eta=\|h\|_\nuc=\|\widehat{h}\|_\nuc.
\end{equation}
Observe that $h^j$ are matrices of rank $\leq 2$, so that $\|h^j\|_\nuc\leq\sqrt{2}H_j$, and since $\widehat{H}=\sum_{j=1}^mh^j$, we have $\|\widehat{h}\|_\nuc\leq \sum_j\|h_j\|_\nuc\leq \sqrt{2}\sum_jH_j$, which combines with (\ref{secondcase_premise}) to imply the first inequality in the following chain:
$$
\begin{array}{l}
k^2\|h\|_{(k)}^2=\|\widehat{h}\|_\nuc^2\leq 2\left(\sum_{j=1}^mH_j\right)^2=2\left(\sum_{j=1}^m[H_j\mu_j^{(q-1)/2}]\mu_j^{(1-q)/2}\right)^2\\
\leq 2\left(\sum_{j=1}^m\mu_j^{q-1}H_j^2\right)\left(\sum_{j=1}^m\mu_j^{1-q}\right) \hbox{\ [Cauchy inequality]}\\
\leq 2q^{-1}D^2\chi(y)[h,h]\left(\sum_{j=1}^m\mu_j^{1-q}\right)\hbox{\ [by (\ref{tthn})]}\\
\leq 2q^{-1}D^2\chi(y)[h,h]\left((m-n)\epsilon^{1-q}+\sum_{j=m-n+1}^m\mu_j^{1-q}\right) \hbox{\ [by (\ref{mujs})]}\\
\leq 2q^{-1}D^2\chi(y)[h,h]\left((m-n)\epsilon^{1-q}+[n^{-1}\sum_{j=m-n+1}^m\mu_j]^{1-q}n\right)\hbox{\ [since $0<q<1$]}\\
\leq 2q^{-1}D^2\chi(y)[h,h]\left((m-n)\epsilon^{1-q}+k^{1-q}n^{q}\right). \hbox{\ [by (\ref{mujs})]}
\end{array}
$$
 Thus, in the case of (\ref{secondcase_premise}) we have
$$
D^2\chi(y)[h,h]\geq {q\over 2}{k^2\over (m-n)\epsilon^{1-q}+k^{1-q}n^{q}}\|h\|_{(k)}^2.
$$
Setting
\begin{equation}\label{alphaeps}
\alpha_\epsilon=q\min[1,{1\over 2}{k^2\over (m-n)\epsilon^{1-q}+k^{1-q}n^{q}}]
\end{equation}
and recalling (\ref{firstcase}), we arrive at the desired inequality (\ref{pound}).
\par\noindent{\bf 7$^0$.} As we have already explained, (\ref{pound}) implies the validity of (\ref{eq111})  with $\alpha_\epsilon$ given by (\ref{alphaeps}). Since
$Y^k_F\subset Y^\epsilon$ and $\alpha_\epsilon\to\beta$ as $\epsilon\to+0$ (see (\ref{beta})), (\ref{target}) follows.
 \qed
\par\noindent{\bf B.} Lemma \ref{LemNN} is the key to the two statements as follows.
\begin{proposition}\label{propNN1} Let $k,m$ be integers such that $1\leq k\leq m/2$, and let $X=\{x\in \bS^m:I\succeq x\succeq0, \Tr(x)=k\}$. The function
\begin{equation}\label{func1}
\begin{array}{c}
\omega(x)={4\over\beta(1+q)}\sum_{j=1}^m|\lambda_j(x)|^{1+q},\\
q=\left\{\begin{array}{ll}\min[1,\ln(k)/\ln(m/k)],&k>1\\
1/(2\ln(m)),&k=1\\
\end{array}\right.,\,\beta=\left\{\begin{array}{ll}1,&k\geq\sqrt{m}\\
q/2,&1<k<\sqrt{m}\\
q/(2\sqrt{e}),&k=1\\
\end{array}\right.\\
\end{array}
\end{equation}
is convex continuously differentiable function on $E$ which is strongly convex, modulus 1 w.r.t. $\|\cdot\|_X$, on $X$ and thus is a d.-g.f. for $X$ compatible with $\|\cdot\|_X$. The $\omega$-radius of $X$
satisfies
\begin{equation}\label{XomegaX}
\Omega_X\leq2\sqrt{{2k\over\beta(1+q)}}.
\end{equation}
\end{proposition}
{\bf Proof.} The only non-evident statement is that $\omega$ is strongly convex, modulus 1 w.r.t. $\|\cdot\|_X$, on $X$, and this is what we are about to prove. Let $\|\cdot\|_{(k)}$ be the norm on $\bS^m$ with the unit ball $Y^k=\{y\in\bS^m:\|\lambda(y)\|_\infty\leq1,\|\lambda(y)\|_1\leq k\}$, and let
$$
\chi(x)={1\over 1+q}\sum_{j=1}^m|\lambda_j(x)|^{1+q}.
$$
When $k\geq\sqrt{m}$, $Y^k$ contains the unit ball of the Frobenius norm, and consequently $\|\cdot\|_{(k)}\leq\|\cdot\|_\Fro$, and $q=1$, meaning that the function $\chi(\cdot)={1\over 2}\|\cdot\|_\Fro^2$ is strongly convex, modulus 1, w.r.t. $\|\cdot\|_\Fro$, and therefore is strongly convex, modulus $\beta:=1$, w.r.t. $\|\cdot\|_{(k)}\leq\|\cdot\|_\Fro$. Let now $k<\sqrt{m}$. In this case $q\in(0,1)$, and therefore, by Lemma \ref{LemNN}, $\chi$ is strongly convex, modulus $\beta:=q\min[1,{1\over 2}k^{1+q}m^{-q}]$, on $Y^k$. Note that $\beta=q/2$ when $k>1$ and $\beta=q/(2\sqrt{\e})$ when $k=1$.
\par
Now observe that $X$ clearly is contained in $Y^k$, implying that $\chi(x)$ is strongly convex, modulus $\beta$ w.r.t. $\|\cdot\|_{(k)}$, on $X$.
At the same time, we claim that the $\|\cdot\|_X$-unit ball $X^s\subset L[X]=\{x\in \bS^m:\Tr(x)=0\}$ contains the set $\{x\in L[X]:\|x\|_{(k)}\leq 1/2\}$, meaning that $\|\cdot\|_X\leq 2\|\cdot\|_{(k)}$ on $L[X]$; as a result, $\chi(\cdot)$ is strongly convex, modulus $\beta/4$ w.r.t. $\|\cdot\|_X$, on $X$, so that $\omega(x)=(4/\beta)\chi(x)$ is strongly convex, modulus $1$ w.r.t. $\|\cdot\|_X$, on $X$, and this is exactly what we want to prove. To support our claim, let $x\in L[X]$ be such that $\|x\|_{(k)}\leq1/2$, and let $x=U\Diag\{\xi\}U^T$ be the eigenvalue decomposition of $x$. Since $x\in L[X]$ and $\|x\|_{(k)}\leq1/2$, we have
$$
(a):\ \sum_{j=1}^m\xi_j=0,\quad (b):\ |\xi_j|\leq 1/2\,\forall j\leq m,\quad (c):\ 2\alpha:=\sum_{j=1}^m|\xi_j|\leq k/2.
$$
Now let us select $\delta_j\geq0$, $1\leq j\leq m$, in such a way that
$$
(d): \ \delta_j\leq 1/2-|\xi_j|\,\forall j,\quad (e):\ \sum_j\delta_j={1\over 2}k-\alpha.
$$
Such a selection is possible due to $|\xi_j|\leq1/2$ (by $(b)$) and $\sum_{j=1}^m[1/2-|\xi_j|]=m/2-2\alpha\geq k/2-\alpha$ (see $(c)$ and take into account that $k\leq m/2$). Now let $\eta^+=2(\xi^++\delta)$, $\eta^-=2(\xi^-+\delta)$, where $\xi^+$ is the vector with coordinates $\max[\xi_i,0]$, and $\xi^-$ is the vector with coordinates $\max[-\xi_i,0]$. We have $\eta^\pm\geq0$ (since $\delta\geq0$) and $\|\eta^\pm\|_\infty\leq 1$ (by $(d)$). Finally, $\sum_j\xi^+_j=\sum_j\xi^-_j=\alpha$ by $(a)$ and by the definition of $\alpha$, whence $\sum_j\eta^+_j=\sum_j\eta^-_j=2\sum_j\delta_j+2\alpha=k$ by $(e)$. These relations imply that the symmetric matrices $x^\pm=U\Diag\{\eta^\pm\}U^T$ belong to $X$, and by construction $x={1\over 2}[x^+-x^-]$, so that $x\in X^s$, as claimed. \qed

\begin{proposition}\label{propNN2} Let $K,M,N$ be positive integers such that $2 K\leq M\leq N$, and let $\|\cdot\|_{(K)}$ be the norm on $\bR^{M\times N}$ with the unit ball $X=\{x\in\bR^{M\times N}:\|\sigma(x)\|_\infty\leq1,\|\sigma(x)\|_1\leq K\}$. Then the function
\begin{equation}\label{dgflast}
\omega(x)={4\over q(1+q)}\sum_{i=1}^M\sigma_i^{1+q}(x),\,q=\min[1,\ln(2K)/\ln(M/K)],
\end{equation}
is convex and continuously differentiable, and its restriction on $X$ is strongly convex, modulus 1 w.r.t. $\|\cdot\|_{(K)}$, on $X$. The $\omega$-radius $\Omega_X$ of $X$ satisfies
\begin{equation}\label{newOmegaX}
\Omega_X\leq2\sqrt{{2K\over q(1+q)}}.
\end{equation}
\end{proposition}
{\bf Proof.} The only nontrivial claim is that $\omega(\cdot)$ is strongly convex, modulus 1, w.r.t. $\|\cdot\|_{(K)}$. When $q=1$, i.e., when $\sqrt{2}K\geq\sqrt{M}$, $X$ clearly contains the ball $\{x:\|x\|_\Fro\leq 1/\sqrt{2}\}$, so that $\|\cdot\|_{(K)}\leq\sqrt{2}\|\cdot\|_\Fro$, and $\omega(x)=2\|x\|_\Fro^2$ is strongly convex, modulus 4, w.r.t. $\|\cdot\|_\Fro$, and thus indeed strongly concave, modulus 2, w.r.t. $\|\cdot\|_{(k)}$. Now let $q<1$. Let $m=M+N$, $n=2M$, $k=2K$, so that $1<k\leq m/2$, and let $\cA(x)=\hbox{\scriptsize$\left[\begin{array}{cc}&x\cr x^T&\cr\end{array}\right]$}$ be the linear embedding of $\bR^{M\times N}$ into $\bS^m$. It is well known that the eigenvalues of $\cA(x)$ are the $n=2M$ reals $\pm\sigma_i(x)$, $1\leq i\leq M$, and $m-n$ zeros. Therefore for the norm $\|\cdot\|_{(k)}$ from Lemma \ref{LemNN} it holds
\begin{equation}\label{thesame}
\|x\|_{(K)}=\|\cA(x)\|_{(k)}\,\forall x\in\bR^{M\times N}.
 \end{equation}
 By Lemma \ref{LemNN}, the function  $\omega^+(y)={2\over q(1+q)}\sum_{j=1}^m|\lambda_j(y)|^{1+q}$ is convex and continuously differentiable  on the entire $\bS^m$, and its restriction on the set $Y=\{y\in \hbox{Im}(\cA):\|y\|_{(k)}\leq 1\}$ is strongly convex, modulus 1 w.r.t. $\|\cdot\|_{(k)}$, on $Y$, implying, due to (\ref{thesame}), that the function $\omega(x)=\omega^+(\cA(x))$ is convex and continuously differentiable on $\bR^{M\times N}$, and its restriction on the unit ball  $X$ of the norm $\|\cdot\|_{(K)}$ is strongly convex, modulus  1 w.r.t. $\|\cdot\|_{(K)}$, on $X$. \qed
\begin{remark}\label{remlast} Note that inspecting the proofs, it is easily seen that the results of Propositions \ref{propNN1}, \ref{propNN2} remain true if when one replaces $\bS^m$ (resp., $\bR^{M\times N}$ with their subspaces comprised of block-diagonal matrices of a given block-diagonal structure. E.g., when $1\leq k\leq m/2$, the function
$$
\omega(x)={4\over\beta(1+q)}\sum_{j=1}^m x_j^{1+q}
$$
with $q,\beta$ given by (\ref{func1}) is a d.-g.f. for the set $X=\{x\in\bR^m: 0\leq x_j\leq 1\,\forall j, \sum_{j=1}^mx_j=k\}$ compatible with the norm $\|\cdot\|_X$ with the unit ball $X^s={1\over 2}[X-X]$ on the space $L[X]=\Lin(X-X)=\{x\in\bR^m:\sum_jx_j=0\}$ (treat $m$-dimensional vectors as diagonals of $m\times m$ diagonal matrices).
\end{remark}
\end{document}